\documentclass{article}
\usepackage{arxiv}

\usepackage{graphicx,amsfonts,here}

\usepackage{lmodern}
\usepackage{amssymb,amsmath}

\usepackage[utf8]{inputenc}
\usepackage[english]{babel}
\usepackage[T1]{fontenc}

\usepackage{siunitx}

\usepackage{ae,aecompl}

\usepackage{csquotes}
\usepackage{url}

\usepackage{subfigure}
\usepackage{subfigmat}

\usepackage{calc}

\usepackage[usenames,dvipsnames]{xcolor}

\usepackage{listings}
\lstdefinestyle{Python}
{
	language=Python,
	basicstyle=\linespread{0.8}\small\ttfamily,
	numbers=right,                   
	backgroundcolor=\color{white},  
	frame=single,                   
	rulecolor=\color{black},        
	tabsize=1,                      
	breaklines=true,                
	breakatwhitespace=false,        
	keywordstyle=\color{Blue},      
	commentstyle=\color{Green},     
	stringstyle=\color{ForestGreen} 
}

\usepackage{nomencl}

\graphicspath{{Graphics/}}

\usepackage[colorlinks=true]{hyperref}

\makenomenclature

\newcommand\Cpp{{C\nolinebreak[4]\hspace{-.05em}\raisebox{.4ex}{\tiny\bf ++ }}}

\title{Advances of the Python--based Fluid--Structure Interaction capabilities included in SU2}

\author{ \href{https://orcid.org/0000-0002-8059-6590}{\includegraphics[scale=0.06]{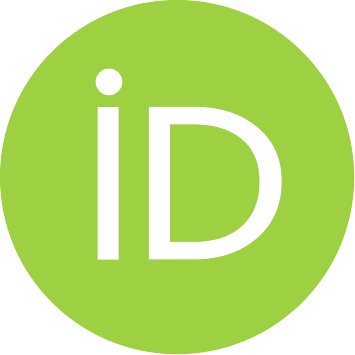}}\hspace{1mm}Nicola Fonzi\\
	PhD student at Department of Aerospace Science and Technology\\
	Politecnico di Milano\\
	Via la Masa 34, Milano \\
	\texttt{nicola.fonzi@polimi.it} \\
	\And
	\href{https://orcid.org/0000-0002-2594-1507}{\includegraphics[scale=0.06]{Graphics/orcid.pdf}}\hspace{1mm}Vittorio Cavalieri \\
	PhD student at Department of Aerospace Science and Technology\\
	Politecnico di Milano\\
	Via la Masa 34, Milano \\
	\And
	\href{https://orcid.org/0000-0003-1651-3532}{\includegraphics[scale=0.06]{Graphics/orcid.pdf}}\hspace{1mm}Alessandro De Gaspari \\
	Assistant Professor at Department of Aerospace Science and Technology\\
	Politecnico di Milano\\
	Via la Masa 34, Milano \\
	\And
	\href{https://orcid.org/0000-0002-4613-1896}{\includegraphics[scale=0.06]{Graphics/orcid.pdf}}\hspace{1mm}Sergio Ricci \\
	Full Professor at Department of Aerospace Science and Technology\\
	Politecnico di Milano\\
	Via la Masa 34, Milano \\
}

\renewcommand{\shorttitle}{Advances of the Python FSI capabilities in SU2}

\hypersetup{
pdftitle={Advances of the Python--based Fluid--Structure Interaction capabilities included in SU2},
pdfauthor={Nicola Fonzi, Vittorio Cavalieri, Alessandro De Gaspari, Sergio Ricci},
pdfkeywords={Aeroelasticity, Python, CFD, Fluid-Structure Interaction, Time-marching, Flutter},
}

\begin{document}

\maketitle

\begin{abstract}

\noindent Current research efforts in aeroelasticity aim at including higher fidelity aerodynamic results into the simulation frameworks. In the present effort, the Python--based Fluid--Structure Interaction framework of the well known SU2 code has been updated and extended to allow for efficient and fully open-source simulations of detailed aeroelastic phenomena. The interface has been standardised for easier inclusion of other external solvers and the comunication scheme between processors revisited. A native solver has been introduced to solve the structural equations coming from a Nastran--like Finite Element Model. The use of high level programming allows to perform simulations with ease and minimum human work. On the other hand, the Computational Fluid Dynamics code of choice has efficient lower level functions that provide a quick turnaround time. Further, the aerodynamic code is currently actively developed and exhibits interesting features for computational aeroelasticity, including an effective means of deforming the mesh. The developed software has been assessed against three different test cases, of increasing complexity. The first test involved the comparison with analytical results for a pitching-plunging airfoil. The second tackled a three-dimensional transonic wing, comparing experimental results. Finally, an entire wind tunnel test, with a flexible half-plane model, has been simulated. In all these tests the code performed well, increasing the confidence that it will be useful for a large range of applications, even in industrial settings. The final goal of the research is to provide with an excellent and free alternative for aeroelastic simulations, that will leverage the use of high-fidelity in the common practise.

\end{abstract}


\printnomenclature

\section{Introduction}

Aeroelasticity has always been an important topic in aerospace engineering. Since the dawn of aviation, aeroelasticity has affected the design of aircraft, as well as developments to avoid flutter \cite{garrick_historical_1981,livne_future_2003}. Recent advances in multi--disciplinary optimization enable to take into account aeroelastic effects in the design process \cite{drela_influence_2019}. This would allow to reduce the need for later redesign and to take advantage of aeroelasticity to design more efficient aircraft \cite{martins_highfidelity_2019}. For example, strategies can be implemented to design longer endurance aircraft and lighter structures, or to install active aeroelastic control systems on--board \cite{stanford_LA_2020,My:AIAAJ:Tesi,My:Scitech2020:AFS}.

However, the common practice to address all these aspects mostly relies on the use of potential aerodynamic methods, namely Doublet Lattice Method (DLM) and Vortex Lattice Method (VLM). Corrections for close to transonic effects must be provided to these methods if that regime is of interest, and this requires more accurate computations \cite{thormann_correction_2014}.

Further, future trends towards highly flexible wings, deep--transonic speeds, and less conventional lifting surface designs, will hinder the application of potential methods. Indeed, Aerodynamic Influence Coefficients Matrices (AIC) are usually computed in a certain configuration and small perturbations are assumed for aeroelastic analysis. This would not hold true for the large displacements exhibited by new designs like the Boeing-787 \cite{wagner_boeing_2009} or High Altitude Long Endurance aircraft \cite{patil_nonlinear_2001}. Further, deeper in the transonic range, where the well known transonic dip occurs \cite{isogai_transonic-dip_1979}, strong nonlinear aerodynamic effects can only be captured with more complex models. 

\noindent Finally, where non conventional configurations are utilised, like T--Tails, it is well known that the basic assumptions behind potential methods cease to be valid \cite{chuban_influence_2005,jennings_effect_1977}. It is interesting to note that newly presented concept planes, all have peculiar characteristics that will hinder the application of classical methods for their analysis \cite{airbus_zero-emission_2021,liebeck_design_2004}.

In all these cases, it is required to have a more complete description of aerodynamics. The most general model that can be used is provided by a Computational Fluid Dynamic (CFD) model. Modern efforts in aeroelasticity are directed towards a better, and more efficient, use of this tool for flutter prediction, and aeroelastic phenomena in general \cite{LI2021106451,cavagna_application_2007}. This can be clearly seen as the latest aeroelastic prediction workshops focused on the use of CFD in this field of research \cite{heeg_plans_2015}.

\noindent In these workshops, the focus was on transonic application, thus conditions where nonlinearities in the flow are present due to the appearance of shock waves. However, nonlinearities can come from an infinite range of sources. There is an excellent review of modern nonlinear aeroelasticity in \cite{dowell_nonlinear_2002}. Some possible sources have been already mentioned above, others, less obvious, may include concentrated nonlinearities in the structure. Indeed, concentrated nonlinearities in control surface attachments are often of interest, and they are usually studied using a quasi-linearisation, obtained via a describing function approach. Usual linear aerodynamic methods are then applied to the resulting model to obtain stability boundaries. However, it has been shown that flutter predictions, obtained via nonlinear time marching simulations, may differ if CFD or potential methods are used \cite{huang_three_2021,ni_flutter_2012}. Thus, it may be of interest to introduce CFD instead of DLM, and assess the stability of these equivalent linear systems.

Therefore, there is the need for always better and easier to use methods for high--fidelity aeroelasticity. An efficient procedure able to couple even complex aerodynamic computations, without being too time--consuming, would help the transition from potential--based methods. Further, it would be desirable to have an open source framework so that custom modifications can be easily embedded.

\noindent A concern specific to industrial applications, is related to the inclusion of new methods in established workflows. The best practises usually involve the use of Nastran as the aeroelastic solver \cite{nastran}, thus it is common to model the structure using this format.

For these reasons, in the current effort, the authors updated and extended the Python--based Fluid--Structure Interaction (FSI) framework available in the open--source code SU2. This framework was originally derived from CUPyDO \cite{thomas_staggered_2017,thomas_cupydo_2019}, to exploit SU2 \cite{palacios_stanford_2013} for the aerodynamic solution, and an external solver for the structural solution. A native structural solver, designed to work with Nastran--like models, has been introduced. The whole framework is written in python and the native solver is as general as possible. With the developed code, high--fidelity aerodynamics can easily be included also in industrial aeroelastic analyses, as the usual structural models will be used. Only one step of the usual workflow will be changed, stepping from the calculation of the aerodynamic forces via DLM to a more accurate calculation. 

\noindent The Nastran--like format is also adopted by the code NeoCASS \cite{cavagna_neocass_2011}, which uses the same input--output scheme of Nastran, but it is fully open--source software. This option may be of great interest for researchers that want to exploit Nastran structural models, without access to the commercial code.

\noindent The FSI interface, which also received many modifications, was kept general, allowing for other structural solvers to be coupled.

As of December 2020, the code has been embedded in SU2 package itself, allowing anyone to exploit these capabilities compiling only one software \cite{SU2_FSI}.

This workflow is efficient enough for a more routinely use of high fidelity aeroelasticity. The developed framework also has the great advantage, compared to other software available, that can be completely based on open source projects. Other programs are already available, and already proved their effectiveness, but most of them are restricted in use. FUN3D, developed by NASA, is a powerful high-fidelity aeroelastic software, based on a fully unstructured formulation, that is however export restricted and only available to US residents \cite{biedron2019fun3d}. EZNSS, which is based on a Chimera structured grid formulation, is also an effective mean of simulating aeroelastic system, but it is restricted to Israeli Air Force \cite{levy2015eznss}. ZEUS, from ZONA Technology, is a high fidelity Euler solver, which is for commercial use only \cite{version_92_2017}. The most common open source choice is OpenFOAM \cite{openfoam}. However, due to the very new community related to SU2 development, some helpful features, especially related to mesh deformation and adjoint studies, are continuously being improved and prove to be extremely helpful in the context of aeroelasticity. Adjoint optimisations have been previously performed using the python interface \cite{bombardieri2021towards}, with a beam model, and the work presented here may be exploited to ease this process. Thus, the authors believe that the present contribution may be of great help for both researchers and practitioners.

The plan of the paper is as follows. In section \ref{sec:methods} of this paper, the used methods are outlined. First, a general explanation about the framework is provided. Next, details are given for the fluid solver, the interface between fluid and structure, and the structural solver. Space will be given to both a general external solid solver and the new native solver.
In section \ref{sec:res}, results obtained with the code are presented. We considered three test cases; a NACA 0012 airfoil, a wing in transonic flow, and a numerical simulation of an entire model in the wind tunnel. The NACA 0012 results are quite common and here reported as a reference to verify the good implementations of the methods. Similar results have already been reported, for SU2, in a previous paper \cite{sanchez_assessment_2016}. However, in the context of the work in \cite{sanchez_assessment_2016}, a specialised structural solver, only able to solve for a typical section airfoil, has been used. Here, we reproduce the same results with the general native structural solver, that can later be used for an arbitrary application. Thus, while the results will be similar in nature, they come from a completely different source. As far as the transonic wing is concerned, this is the Benchmark SuperCritical Wing (BSCW). Again, forced oscillations have already been studied for this case. However, in this paper, we also present flutter analysis. The full power of the implementation will be proved in the last set of results where a complex structural model, fully flexible (the BSCW only has pitch and plunge degrees of freedom), and immersed in a strongly nonlinear flow field, will be studied in terms of dynamics and stability.
Finally, in section \ref{sec:conclusions}, conclusions will be drawn and future work directions outlined.

\section{Framework for aeroelastic computation}\label{sec:methods}

The overall goal of this research is twofold. First, the update of the SU2 Python--based FSI framework, an environment where the well known CFD solver SU2 is coupled with different structural solvers. Second, the introduction of a native solver for the integration of structural equations coming from a Nastran--like model.

\noindent The reader is encouraged to read the original paper where CUPyDO is presented \cite{thomas_cupydo_2019}, as this work is a natural continuation of it.

\begin{figure}[htp]
  \centering
  \includegraphics[width=1\textwidth]{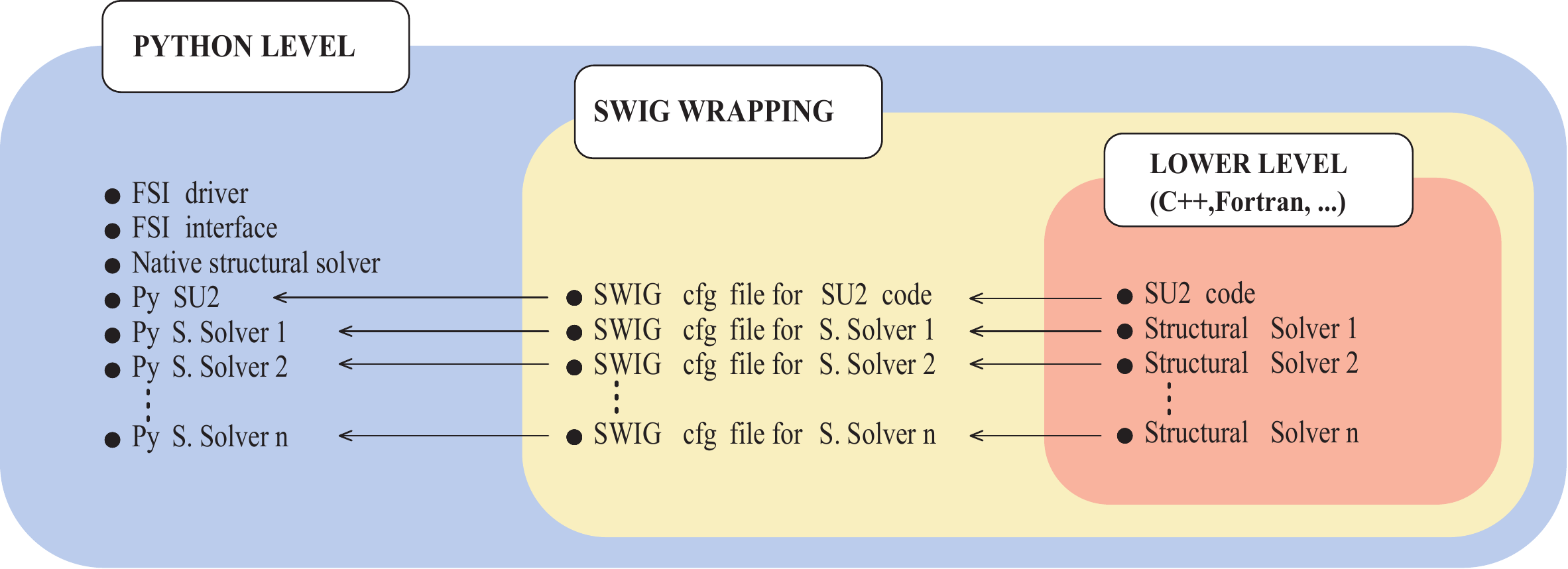}
  \caption{Different code levels of the Python--based FSI framework.}
  \label{fig:Levels}
\end{figure}

The current state of the framework is presented in \ref{fig:Levels} and a typical call of the main driver is summarised in the code snipped below.
\newpage
\begin{lstlisting}[style=Python,showstringspaces=false]
# This code snippet is taken from the FSI driver

# This is the fluid solver (e.g SU2 wrapped in python)
import pysu2
# This contains the interface and the configuration parser
import FSI
# Import also mpi
from mpi4py import MPI
comm = MPI.COMM_WORLD

# Parse the configuration file
FSI_config = FSI.FSIConfig(configFile)

# Initialise the fluid solver
FluidSolver = pysu2.CSinglezoneDriver(FSI_config["CFD_CONFIG"],comm)

# Initialise the solid solver
if FSI_config["CSD_Solver"]=="NATIVE":
    from SU2_Nastran import pysu2_nastran
    SolidSolver = pysu2_nastran.Solver(FSI_config["CSD_CONFIG"])
elif FSI_config["CSD_Solver"]=="EXTERNAL1":
    import pysu2_external1
    SolidSolver = pysu2_external1.Solver(FSI_config["CSD_CONFIG"])
...
else:
    raise Exception("Unrecognised solver")

# Initialise the interface
FSIInterface = FSI.Interface(FSI_config, FluidSolver, SolidSolver)

# Create the interpolation
FSIInterface.interfaceMapping(FluidSolver, SolidSolver, FSI_config)

# Launch the simulation
# FSI_config contains the flag to set an imposed motion
# In that case, the solution process will be different
if FSI_config["TIME_MARCHING"] == "YES":
    FSIInterface.UnsteadyFSI(FSI_config, FluidSolver, SolidSolver)
else:
    FSIInterface.SteadyFSI(FSI_config, FluidSolver, SolidSolver)

\end{lstlisting}

It can be noted that support for different structural solvers is still allowed, but there is also the new native solver, which is totally written in python, and integrated in SU2 itself. The wrapping process is the same that has been previously used. SWIG \cite{beazley1996swig} is used to obtain a Python Application Programming Interface (API) to the lower level functions. Obviously, not all the lower level functions have a counterpart in python, but the main ones, required to drive the simulation, are mapped. The interface has been standardised so that, when adding a new structural solver, no modifications are required. The SWIG configuration file must be written so that the resulting python object, wrapping a general structural solver, contains a specific set of functions. These functions are required to ask the positions of the structural nodes, ask if these nodes are halo (in case parallel computing is exploited), get the indices of these nodes, apply loads onto them, update their solution in time, and write the solution to file. More than that, three main functions, namely the function to run a step of the integration and the constructor--destructor, are required. More technical details are available in the documentation.

\noindent If all these required routines are present, regardless of the process underlying in the lower level functions, the coupled simulation will run smoothly.

\begin{figure}[htp]
  \centering
  \includegraphics[width=0.32\textwidth]{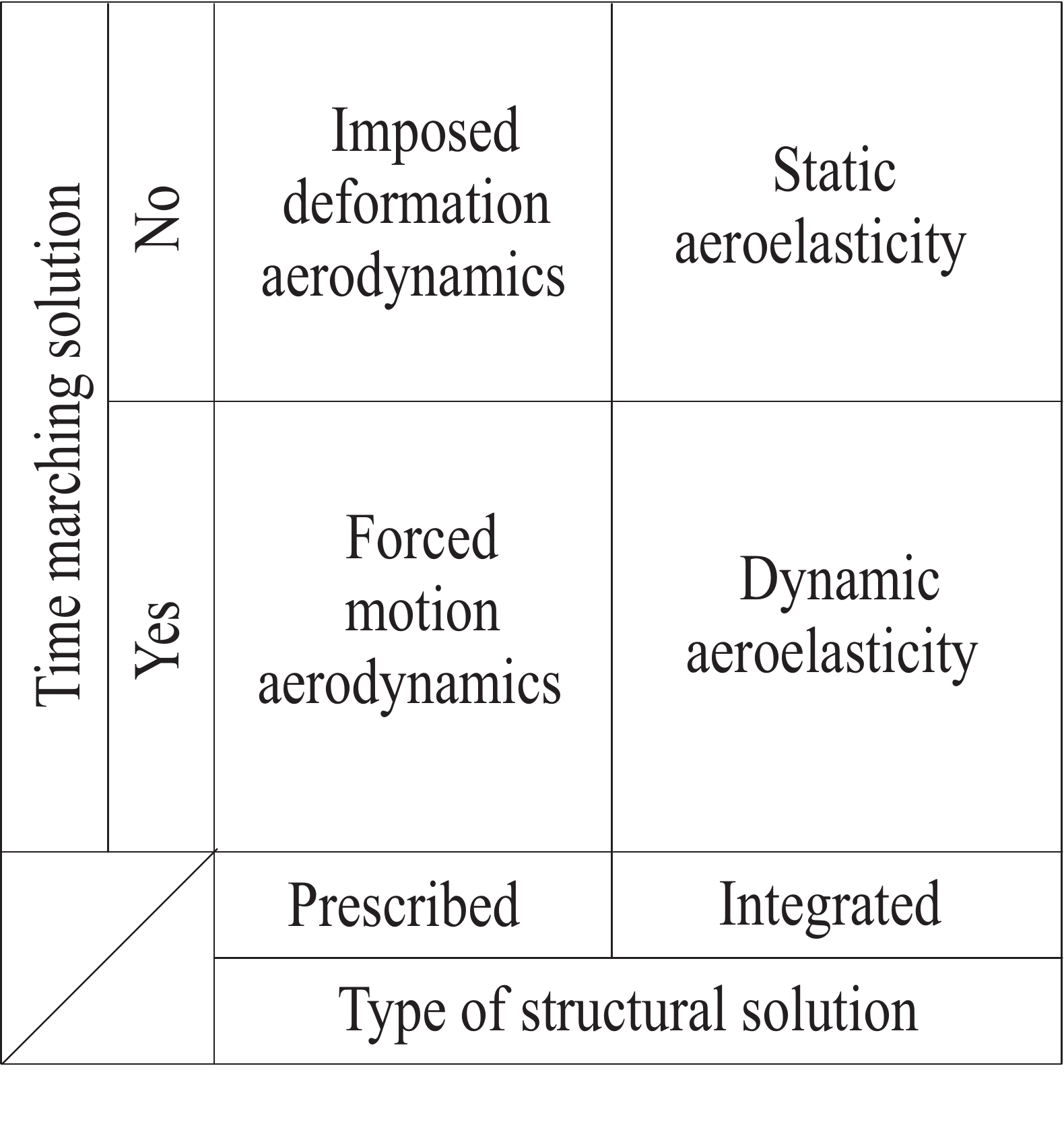}
  \caption{Available simulation settings for the current framework}
  \label{tab:TablePossibleSimulations}
\end{figure}

\noindent The solution process has also been generalised and standardised so that four kinds of simulations are now allowed, as depicted in Fig. \ref{tab:TablePossibleSimulations}.

In case the structural motion is prescribed, the aerodynamic response to specific inputs is computed. Here, the structural model is actually only used to provide the structural node positions required for the interface script.

\noindent Computing the aerodynamic responses is really useful if, for example, the aerodynamic system must be identified \cite{fonzi2020data}. These responses can be either computed in a steady or unsteady simulation. In the former case, we may want, for example, to understand how, with a certain displacement field, the aerodynamics changes around our body. Using the present code it will not be required to prepare again the mesh, as the framework will automatically deform it around the new structural interface. An ongoing research inside the Department of Aerospace Science and Technology at Politecnico di Milano is, for example, exploiting this feature for shape optimisation of morphing airfoils.

On the other hand, if the structural solution is to be obtained from the structural solver after an actual integration of the equations of motion, the aeroelastic system will be studied. In this case, the coupling between the fluid solver and the structural solver is provided in a tightly coupled framework.

In particular, for each time step, or just once in case of a steady computation, the solution process is the following:

\begin{enumerate}
    \item The initial positions of the structural nodes are obtained and interpolated via Radial Basis Functions (RBF) onto the aerodynamic mesh. In this phase, if required, grid velocities are computed. Please note that, if this is the very first time step and there is an initial deformation, fictitious velocities are avoided by imposing zero grid velocity in all the domain.
    \item The fluid solver is run, obtaining the distribution of aerodynamic forces on the surface at the interface between fluid and structure. SU2 is a vertex centered finite volume solver, thus the solution is already obtained at the aerodynamic nodes.
    \item If the structural solution is prescribed, the displacement field is imposed.
    \item Else:
    \begin{itemize}        \item The forces are interpolated via RBF onto the structural nodes.
        \item The structural solver is run to obtain the new positions.
    \end{itemize}
    \item Convergence is checked. This is done computing the root mean square of the incremental structural displacements, between the previous FSI iteration and the current one.
    \item If convergence is reached, the structural displacements for the next time step are predicted.
    \item Else, the structural solution is relaxed with an Aitken relaxation, and the steps of this list are repeated.
\end{enumerate}

\begin{figure}[ht]
\centering
\includegraphics[width=0.75\textwidth]{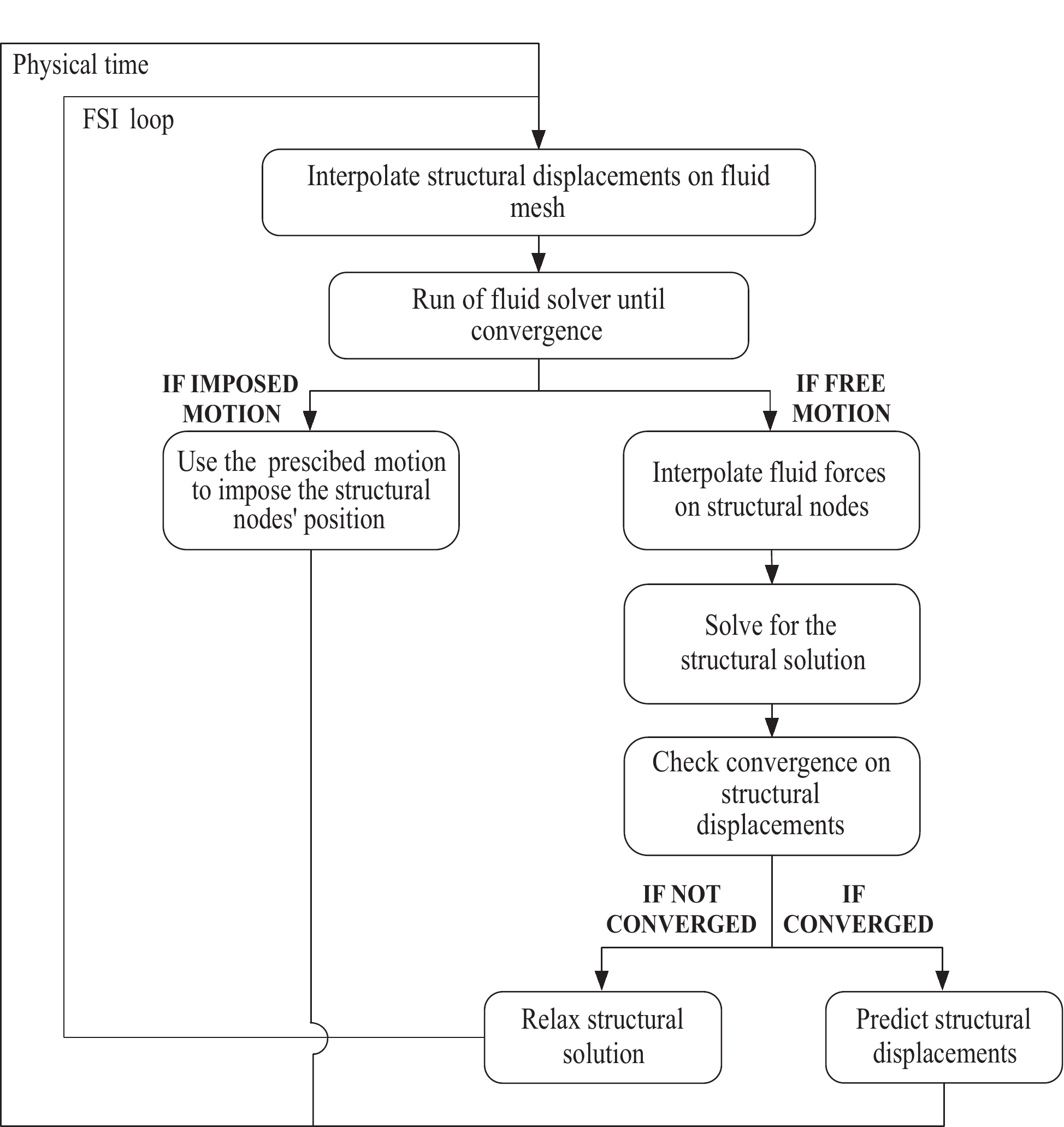}
\caption{Solution process flow chart.}
\label{fig:solution_flow}
\end{figure}

\noindent The process outlined above can also be graphically seen in Fig. \ref{fig:solution_flow}.

\noindent It must be noted that, if the structural solver used does not allow for one of the outlined solution processes, the interface will still run with the remaining options. So, for example, if the structural solver of choice does not allow to impose a certain motion, but can only solve for the nodes' position, the interface will still be able to run the steady and unsteady aeroelastic simulations, without raising errors. This is useful as it increases the flexibility of the code.

In the next subsections, more details will be provided about the different components of the framework.

\subsection{Aerodynamic solver - SU2}

The open source code SU2 was initially developed at Stanford University \cite{palacios_stanford_2013}. The solution method uses a finite volume formulation, vertex centred, with a dual mesh grid. This formulation may offer advantages in terms of accuracy and stencil size, depending on the problem at hand \cite{diskin_comparison_2009}. The entire code has been developed from the very beginning with attention to maximum flexibility, allowing for easy inclusion of new routines and thus making it the perfect platform for multiphysics problems \cite{economon_su2_2016}. 

\noindent The code has demonstrated good capabilities of handling FSI problems in the past \cite{sanchez_assessment_2016} and the authors believe that, in the future, it may became an important resource for practitioners.

\noindent Further, the \Cpp code of SU2 offers the possibility to be wrapped in a Python API, so that high level functions can be called with ease. Indeed, the CFD solver can be called as a normal Python module, giving the opportunity to easily create scripts for optimisation or complex coupled simulations, importing also a structural solver that allows similar flexibility.

\noindent SU2 also includes an internal FEM solver, which is used for multiphysics simulation. However, this solver has its own format and may not be suited for all the possible user's requirements. Thus, the need for coupling with external solvers.

Several numerical schemes are available. In the context of this work, the Jameson-Schmidt-Turkel scheme has been utilised for the convective part of the equations \cite{jameson_time_1991,jameson_origins_2017}. The choice is due to the great range of problems that this method can tackle, including transonic problems. Green-Gauss reconstruction schemes are used for the viscous part.

\noindent Both steady and unsteady simulations will be presented. In all cases, the pseudo time has been used to converge the solution with a local Courant-Friedrichs-Lewy (CFL) number that will be specified in the relevant section of the article. At each pseudo time iteration the linear solver provides the change in solution. Again, different linear solvers are available and the chosen one will be specified on a case by case basis. If unsteady simulations are performed, finite difference approximations, second order in time, are used for the time derivative term.

\noindent In the context of FSI problems, new positions of the grid nodes can be communicated to the CFD solver via the Python wrapped functions. The fluid solver wrapping has now been modified so that the FEM solver internal to SU2 itself is used as a mesh solver. This allows to use all the options that the internal solver has to improve the efficiency and effectiveness of the mesh deformation step. Grid velocities for the Arbitrary Lagrangian-Eulerian (ALE) method are then computed using finite differences in time.

\subsection{Aero--structure interface method}

The interpolation between fluid and structure is performed within the Python script via a RBF interpolation \cite{de_boer_mesh_2007}. 

\noindent The displacements at the structure side are obtained starting from the positions of the undeformed structural nodes and the obtained fluid mesh displacements are also defined starting from the initial mesh.

\noindent The RBF method involves the choice of a kernel function and a polynomial order for the interpolation. In this case, the interpolation polynomial is linear, as this recovers correctly rigid body translations \cite{beckert_multivariate_2001}, and the kernel function is the CP C2 \cite{de_boer_mesh_2007}.

The interface is completely parallelized, in order to obtain maximum performance. In the older version, mpi4py \cite{dalcin2008mpi} was used as a means of communication between processors, and petsc4py \cite{dalcin2011parallel} for the solution of the linear problem given by the RBF interpolation. The same approach is kept here, but all the transfers between processors are done using numpy arrays, as, in this case, mpi4py provides maximum transfer speeds. 

\subsection{Generic structural solver}

In the previous sections it was explained how the interface is standardised to accept any structural solver, provided that it has the required wrapped functions.

\noindent The only modification required to the code, will be the inclusion of the new module in the list of accepted structural solver. In this way, the FSI driver will recognise the module and use it in the FSI interface.

Unfortunately, due to the continuous development of SU2, the support for some structural solvers that were wrapped in the context of the old version was lost. For this reason, the authors took the chance to standardise the requirements for the structural python module, and start again wrapping different solvers.

\noindent As of today, efforts are directed in the creation of a module for the commercial code Abaqus, and others are planned for the near future.

\subsection{Native structural solver}

The native structural solver offers an efficient and straightforward means to obtain an aeroelastic solution using SU2 as aerodynamic solver. The preparation of the inputs for the structural solver usually works in the following way, but it is not limited to this. First, a modal analysis is performed on the structural model with the user's favorite choice of FEM solver. Next, the results have to be translated in Nastran--like format. Of course if Nastran, or NeoCASS \cite{cavagna_neocass_2011,fonte_neocass_ifasd}, which supports the same Nastran format, is directly used, no further translations are required. 

\noindent Two sets of information are required. The first, is the set of grid nodes’ locations, which are required for the interpolation between fluid and structure. The second must contain the modal shapes normalised for unit mass and the modal frequencies. The structural solver will use the latter to build the system of equations to be solved, and the former to define the displacements, velocities, and accelerations of the physical structural nodes, as a function of the modal solution. The physical variables are required due to the coupling with the aerodynamic solver.

\noindent It must be noted that the structural solver is usually used with the above preparation scheme, but not limited to this. Indeed, there are specific keywords to provide it with a non-diagonal system of modal equations, which are, for example, obtained with the use of fictitious masses \cite{Fictitious_Mass_Aeroelastic2:Karpel,Fictitious_Mass_Aeroelastic3:Karpel} or adding static modes. Further, similar keywords are also available if a different scaling than the unit mass is to be used.

\noindent Usually, the structural equations solve for the dynamics of a set of modes of the structure, but an expert user may introduce an arbitrary system of equations to be integrated, provided that it is given to the solver in the appropriate way. So, for example, we may directly solve for all the degrees of freedom of a structural system, using the identity matrix instead of the modal matrix. Overall, the native structural solver was born as a fast means to couple Nastran modal analyses with SU2, but can be used with much more flexibility to integrate any system of structural equations.

\begin{figure}[htp]
  \centering
  \includegraphics[width=0.5\textwidth]{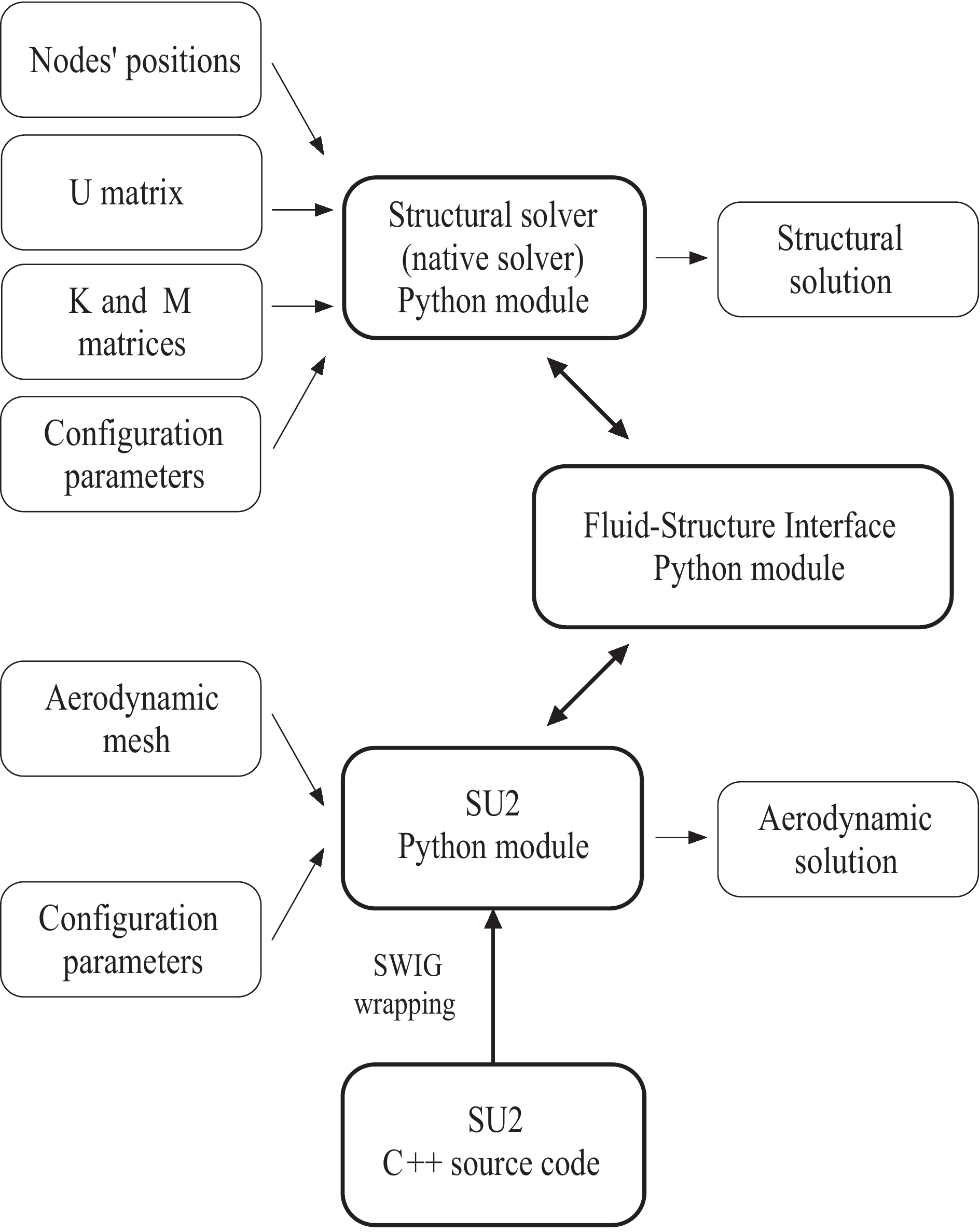}
  \caption{Framework layout in case the native solver is used.}
  \label{fig:framework}
\end{figure}

\noindent In case the native solver is used, the input-output relations of the framework are as represented in Fig. \ref{fig:framework}. In the figure, thin lines are used to identify inputs or outputs, while heavier lines are used to mark the modules. The $\mathbf{U}$ matrix contains the relation between the amplitude of the degrees of freedom and the displacements of all the structural nodes, which may be the matrix of mode shapes, in case modal coordinates are used.

In the particular case of this work, the results of an eigenanalysis performed in Nastran are used for the solution of the aeroelastic problem in modal coordinates, as these already provide the correct format for the inputs to the structural solver.

The equations, in case of decoupled modal equations normalised for unit mass, are stated as:
$$
\begin{cases}
  \begin{array}{ll}
    \ddot{q}_1 + \omega_1^2 q = \tilde{F}_1 \\
    \ddot{q}_2 + \omega_2^2 q = \tilde{F}_2 \\
    \dots \\
    \ddot{q}_n + \omega_n^2 q = \tilde{F}_n \\
  \end{array}
\end{cases}
$$
\nomenclature{$q_i$}{Generalised modal coordinate}
\nomenclature{$\omega_i$}{Pulsation of the i-th mode}
\nomenclature{$\tilde{F}_i$}{Generalised force on the i-th mode}
\nomenclature{$\mathbf{U}$}{Matrix mapping the degrees of freedom to the physical structural nodes}

\noindent Where $q_i$ is the generalised modal coordinate, $\omega_i$ is the related pulsation, $F_i$ it generalised force, and $n$ is the total number of modes considered. The generalised force, in turn, is obtained using the mode shapes matrix $\mathbf{U}$ as: $\tilde{F}_i = U_i^T \mathbf{F}$, with $\mathbf{F}$ the vector of structural nodal forces.

\noindent The structural solver offers the opportunity to include a proportional damping. This is specified as a percentage of the critical damping and adds to each equation a term $2\xi_i\omega_i\dot{q}_i$. Where $\xi_i$ is the percentage of structural modal damping given as an input to the solver.
\nomenclature{$\xi_i$}{Modal damping}

\noindent If the option for non-decoupled equations is used, the introduction of damping is not so straightforward. In typical applications, the same modal damping is prescribed in a range of frequencies. Thus, it is assumed that a certain $\xi$ is provided to the solver and this must be applied to all the modes resulting from the eigenanalysis of the nondiagonal matrices. The modal damping is thus applied in the following way. First, an eigenanalysis is performed obtaining the eigenvalues $d_i$ and eigenvectors $\mathbf{V}_i$. The mass matrix is then diagonalised as:

$$\mathbf{\tilde{M}} = \mathbf{V}^T \mathbf{M} \mathbf{V}$$
\noindent The diagonal damping matrix elements can then be obtained as:
$$\tilde{C}_i = 2 \xi_i d_i \tilde{M}_{ii}$$
\noindent Finally, the obtained matrix is brought back to the initial mathematical space as:

$$\mathbf{C} = \mathbf{V}^{-T} \mathbf{\tilde{C}} \mathbf{V}^{-1}$$
\nomenclature{$\mathbf{M}$}{Mass matrix}
\nomenclature{$\mathbf{V}$}{Matrix of the eigenvectors of the mass matrix}
\nomenclature{$\mathbf{C}$}{Damping matrix}

At each time step, the modal equations are advanced in time with a generalised-$\alpha$ algorithm \cite{chung_time_1993}. This avoids the modification of the system into a set of first order Ordinary Differential Equations (ODEs) and can also provide excellent numerical properties, like unconditional stability and no numerical damping of dynamics with frequencies below half of the sampling frequency.

\noindent In case of a steady solution, the inertia forces are anyway taken into account. In this way, also in the static computation the Newmark method is used, instead of a normal linear solver, adding numerical stability to the coupled simulation.

\section{Validation test cases}\label{sec:res}

Three test cases are here presented, with the purpose of validating the framework for different aerodynamic conditions and applying it to structural models of increasing complexity. All the additions and modifications will be tested, thus the native solver will be used with all the possible simulation types presented in \ref{tab:TablePossibleSimulations}.

\subsection{NACA 0012 airfoil}

The first test case here introduced concerns a NACA 0012 airfoil. The ultimate goal of this section is to demonstrate the capability of the proposed FSI tool in solving a typical basic aeroelastic problem, namely the flutter of a two-dimensional pitching-plunging airfoil.
This application is handled step by step, following the schematic presented in Fig.~\ref{tab:TablePossibleSimulations}. In this way, the features of the code and the available solution methods are gradually shown and validated.

The reference model is a pitching-plunging NACA 0012 airfoil in a free-stream flow.
The structural model is made by a single point, positioned at the rotation axis, with two degrees of freedom, pitch and plunge. 
Inertia and mass of the airfoil are concentrated at the center of mass of the airfoil, at a certain distance from the rotation axis. The equations of motions are available
analytically and read:

$$
\begin{cases}
  \begin{array}{ll}
    m\ddot{h} + S_m\ddot{\theta} + C_{h}\dot{h} + K_{h}h = -L \\
    S_m\ddot{h} + I\ddot{\theta} + C_{\alpha}\dot{\theta} + K_{\alpha}(\alpha+\theta) = \mathcal{M} \\
  \end{array}
\end{cases}
$$

\noindent Where $m$ is the mass of the airfoil, $I$ the inertia around the rotation axis, $S_m$ the static moment of inertia at the rotation axis, $C$ and $K$ the damping and stiffness respectively. $L$ and $\mathcal{M}$ are the lift and pitching up moment. The degrees of freedom are $\theta$, the pitch, and $h$, plunge. On the other hand, $\alpha$ is the geometric angle of attack.

\nomenclature{$m$}{Mass}
\nomenclature{$I$}{Inertia}
\nomenclature{$S_m$}{Static moment of inertia}
\nomenclature{$K_h$}{Stiffness for the plunging mode}
\nomenclature{$K_\alpha$}{Stiffness for the pitching mode}
\nomenclature{$C_h$}{Damping for the plunging mode}
\nomenclature{$C_\alpha$}{Damping for the pitching mode}
\nomenclature{$\alpha$}{Geometric angle of attack}
\nomenclature{$\theta$}{Pitch angle}
\nomenclature{$h$}{Plunge displacement}
\nomenclature{$L$}{Lift}
\nomenclature{$\mathcal{M}$}{Pitching moment}

\noindent However, as we want to develop a general framework, we will not use these equations, but we will rather build a really simple FEM model to represent the airfoil. The mesh is extremely simple; it is realised using a set of rigid elements that connects several slave nodes to a single master node, positioned on the rotation axis. The master node only has two degrees of freedom: pitch and plunge.
One of the slave nodes, at the position of the center of mass, houses the mass and inertia of the airfoil.

\begin{figure}[htp]
\centering
\includegraphics[width=0.75\textwidth]{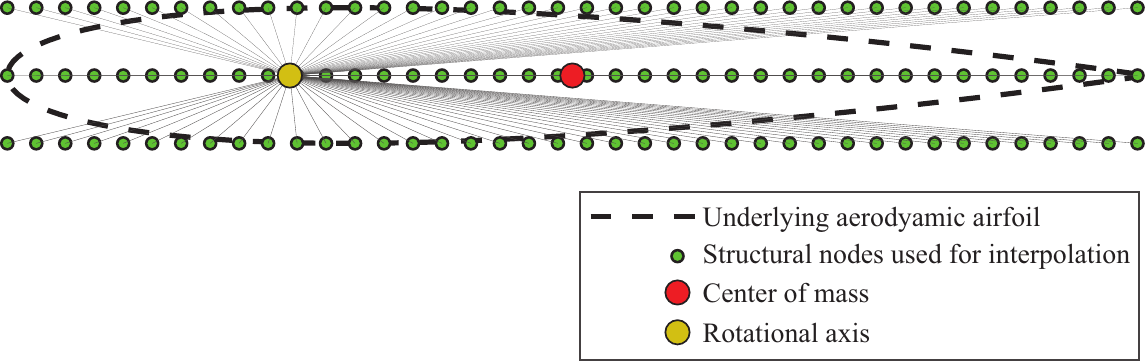}
\caption{FEM mesh for the NACA 0012 airfoil.}
\label{fig:nastran_model}
\end{figure}

The interpolation between the structural and fluid meshes uses RBF. The limitation of this interpolation is due to the fact that, if a 2D problem is concerned, the structural points cannot all lie on the same line. Equivalently, if a 3D problem is tackled, the points cannot lie all in the same plane. For this reason, the thickness is represented in the FEM structural mesh as shown in Fig.~\ref{fig:nastran_model}.

\noindent It can be noted that the exact geometry of the airfoil is not required. The interpolation will take care of displacing correctly the fluid mesh. However, thickness must somehow be represented.

In what follows, the results of four different analyses of increasing complexity are reported, validating the proposed methods on the test case at hand.

\noindent For the first three subsections, the aerodynamic model will be based on Reynolds Averaged Navier-Stokes (RANS) equations, with Menter's Shear Stress Transport (SST) turbulence model, and a pseudo-time CFL number of 20. In the last subsection, related to the dynamic aeroelasticity, also a Euler model will be used, again with a local CFL of 20. In both cases, the linear solver is a Flexible Generalised Minimum Residual. This solver is relatively memory intense, but, for this small case, provides the best performance in terms of accuracy.

\begin{figure}[ht]
\centering
\includegraphics[width=0.6\textwidth]{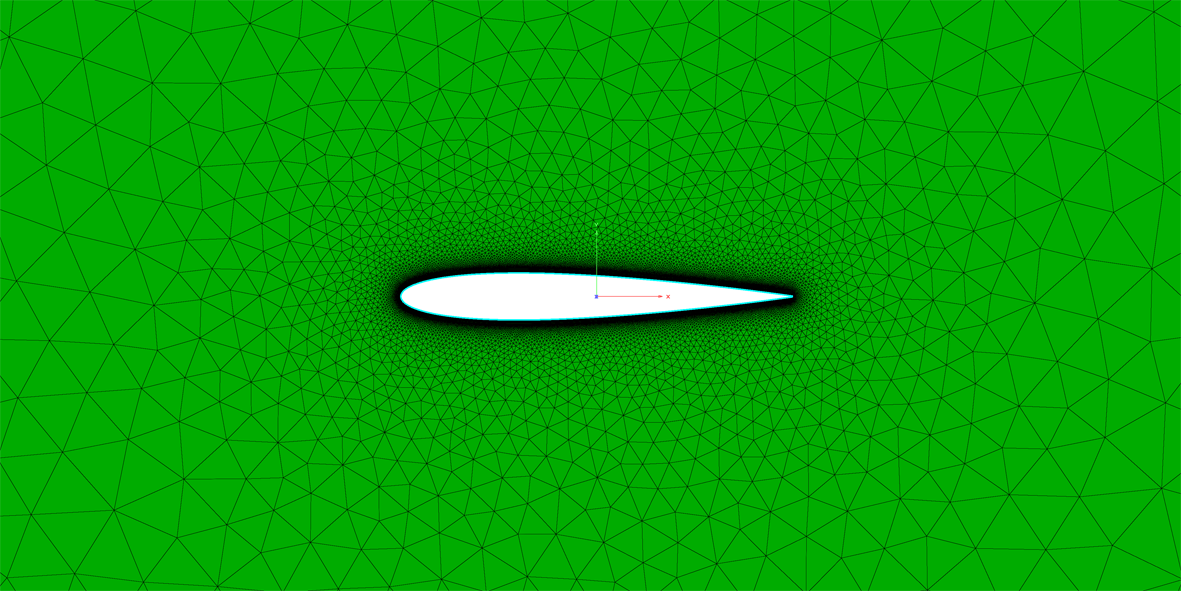}
\caption{Aerodynamic mesh for the RANS simulations of the NACA 0012.}
\label{fig:CFD_Mesh}
\end{figure}

\begin{figure}[ht]
\centering
\includegraphics[height=0.28\textwidth]{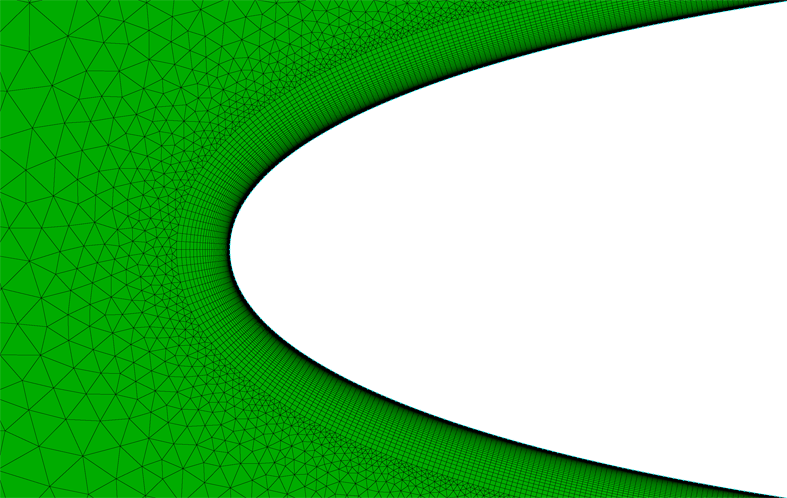} \quad
\includegraphics[height=0.28\textwidth]{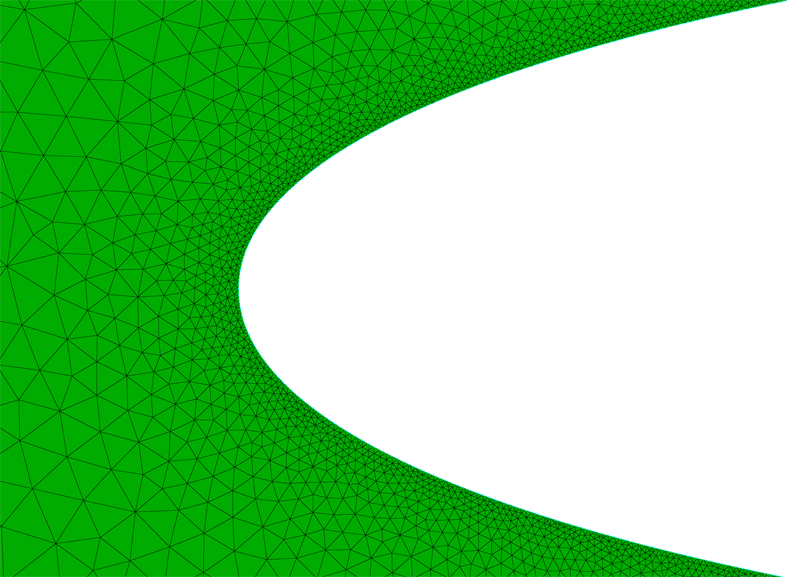}
\caption{Different treatment of the near wall cells in RANS and Euler simulations, for the NACA 0012 airfoil.}
\label{fig:CFD_Mesh_zoom}
\end{figure}

\noindent Depending on the aerodynamic model, two different meshes are used. In the first, for the RANS simulations, the fluid domain is discretised with 133,000 nodes, with refining close to the airfoil surface, in order to correctly represent the turbulent boundary layer. The first cell is placed at a height of $y^+\approx 1$. In the second mesh, there is no need to resolve the boundary layer. The distribution of points, away from the airfoil surface, is the same as before. A total of 19,500 nodes are used in this case. The base mesh for both models is pictured in Fig.~\ref{fig:CFD_Mesh}, while Fig.~\ref{fig:CFD_Mesh_zoom} compares the different treatments of the layers close to the surface.

\subsubsection{Imposed deformation aerodynamics}\label{subsec:naca_steady}

\begin{figure}[ht]
\centering
\includegraphics[width=0.65\textwidth]{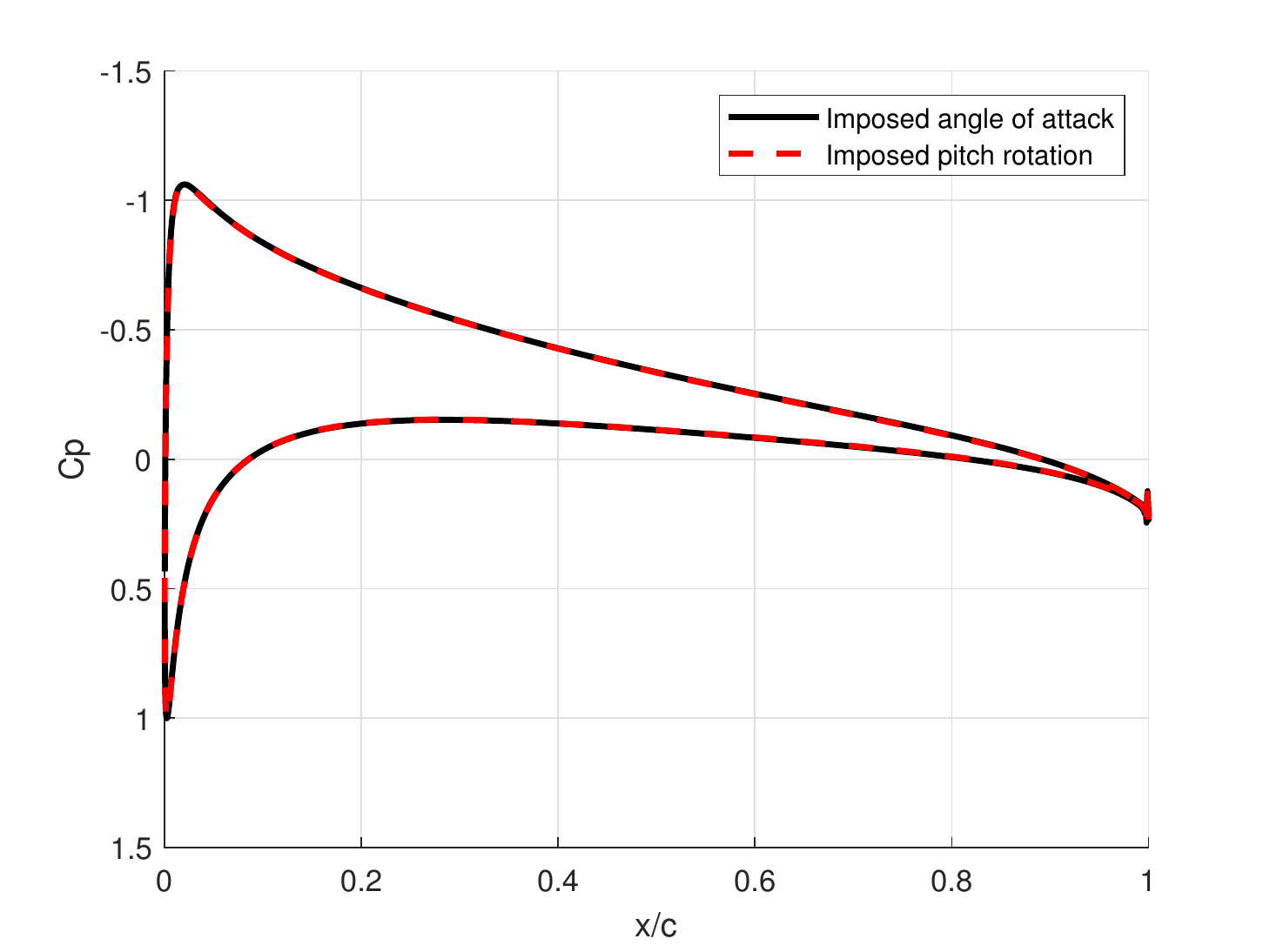}
\caption{Coefficient of pressure distribution around the NACA 0012 airfoil, at $3^{\circ}$ of angle of attack, as computed imposing the flow direction in SU2 or the pitch rotation in the FSI interface.}
\label{fig:naca_steady_cp}
\end{figure}

At the beginning, a simple steady aerodynamic analysis is examined to validate the capability of the interface to impose mesh deformation.
The analysis at an angle of attack of 3 degrees is repeated twice.
First, an angle of attack of 3 degrees is directly set in SU2 in the free-stream definition. Later, an angle of attack of 0 degrees is assumed, but a pitch rotation of 3 degrees is imposed through the fluid-structure interface.
The aerodynamic conditions for this case are representative of a high Reynolds (Re) number and low Mach number flow. The Re number is equal to 6 millions, while the Mach number is equal to 0.1. The temperature is set to $\SI{293.15}{K}$. Given these conditions, the flow will not behave much differently from what can be predicted via common linear incompressible models.
The simulation is considered converged when the root mean square of the residuals of the density equation is decreased of 5 orders of magnitude.
The comparison of the pressure coefficient distribution obtained in the two cases, reported in Fig.~\ref{fig:naca_steady_cp}, does not exhibit any difference, demonstrating the ability of the interpolation.

\subsubsection{Static aeroelasticity}

In the second problem here considered, the two structural degrees of freedom of pitch and plunge are free to change, but the steady equations for the flow are solved. It should be noted that, equivalently to the solution of the aerodynamics which exploits the pseudo time integration, the structural equations are also solved integrating in time. Indeed, including damping and inertia in the structural solution helps convergence as it avoids over/under shoots of the solution. In this case, the time step chosen is usually relatively large, as there are no constraints given by the Nyquist criteria, or the numerical damping. The inertial and damping characteristics can also be chosen arbitrarily, as, in the limit of convergence, both velocity and acceleration are zero. The stiffness in pitch and plunge are respectively $K_h=\SI{205}{N/m}$ and $K_\alpha=\SI{2025}{N.m/rad}$.
The aerodynamic conditions are coincident with those used for the previous test case, except for the Mach number, which is reduced in order to maintain small deformations. The Mach number is halved. The initial angle of attack of the airfoil is also the same, thus it is $\alpha = 3^\circ$. The convergence criteria still requires the density residuals to drop of 5 orders of magnitude.
As far as the FSI loop is concerned, convergence is based on the difference of the norm of the displacements vector between two FSI iterations, as shown in Fig. \ref{fig:solution_flow}, which has to be smaller than 1 micrometer.
The result of the SU2 analysis is compared with theoretical solutions, and with the X-FOIL software. The former is obtained from simple linear incompressible theories.
As the rotation axis is placed exactly on the quarter-chord, and the airfoil is symmetric, no pitch rotation is predicted by the theory. As far as the plunge is concerned, the equation will simply read:

\begin{equation}
    K_h h = \frac{1}{2} \rho U_\infty^2 S C_{l_\alpha} \alpha
\end{equation}

\noindent Solving for $h$, and using the values for the present example, we obtain:

\begin{equation}
    h = \frac{1}{2} \rho U_\infty^2 S \frac{C_{l_\alpha}}{K_h} \alpha = \SI{0.29}{m}
\end{equation}

\nomenclature{$\rho$}{Fluid density}
\nomenclature{$U_\infty$}{Free-stream velocity}
\nomenclature{$S$}{Reference surface}
\nomenclature{$C_{l_\alpha}$}{Slope of the lift curve}

\noindent In order to also capture the effect of the pitching moment, results from X-FOIL, in terms of coefficient of lift and coefficient of moment, have been used \cite{drela_xfoil_1989}, with similar Re.

In Table \ref{tab:naca_static} there is a comparison of the results given with the three different methods, showing good match in all cases.
The values predicted by SU2 show almost no pitch rotation and a similar plunge to the other methods.

\begin{table}[htp]
\caption{Pitch and plunge displacements, as predicted by various methods.}
\label{tab:naca_static}
\centering
\begin{tabular}{@{}ccc}
\hline
Model			    & $h$ (\SI{}{m})    		& $\theta$ (\SI{}{rad})	\cr
\hline
Theory 				&	0.289		&	0			\cr
X-Foil				&	0.281		&	4.0e-04		\cr
SU2					&	0.306		&	3.4e-4		\cr
\hline
\end{tabular}
\end{table}

\subsubsection{Forced motion aerodynamics}

\begin{figure}[ht]
\centering
\includegraphics[width=0.7\textwidth]{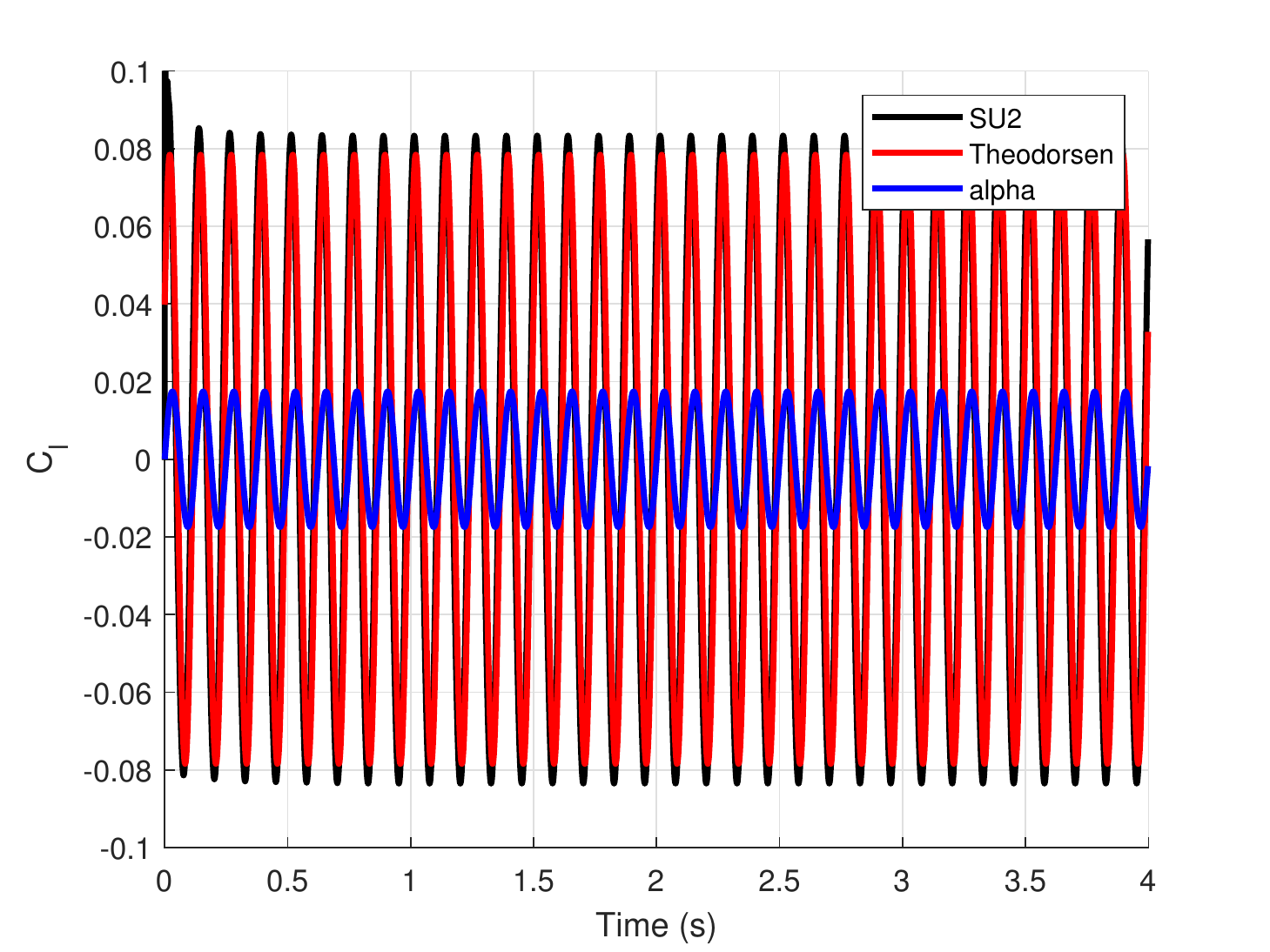}
\caption{Unsteady coefficient of lift of the NACA 0012 airfoil, computed by SU2 with the FSI interface and by the Theodorsen function. Imposed sinusoidal pitching motion with zero mean angle of attack, $1^{\circ}$ amplitude, and $\SI{8}{Hz}$ frequency.}
\label{fig:naca_forced_cl}
\end{figure}

The first time-variant analysis here considered is the response to a forced motion. The plunge displacement is fixed while the pitch angle is imposed as a sinusoidal wave with an amplitude of 1 degree, no bias, and a frequency of $\SI{8}{Hz}$.
This prescribed motion is imposed through the interface. The equations of motion for the structure are not integrated. The equations of fluid dynamics are solved considering at each time step the shape associated to the assigned evolution of the pitch angle.
The simulation is run in SU2 with a Mach number of 0.1, and Re number equal to 6 millions. Each time step, with size of 1 millisecond, the aerodynamic convergence is reached with a drop of around 5 orders of magnitudes in the density residuals.
After a small initial transient, the solution reaches a periodic oscillation.
The comparison of the lift coefficient from the CFD simulation with the one from Theodorsen theory is presented in Fig.~\ref{fig:naca_forced_cl}.

\noindent The theory predicts a lift coefficient with the following expression:

\begin{equation}
    C_l = 2\pi \frac{c}{4} \bigg[\frac{\dot{\theta}}{U}-(x_f-c/2)\frac{\ddot{\theta}}{U^2}\bigg] + 2\pi C(k)\bigg[\alpha+\Big(\frac{3c}{4}-x_f\Big)\frac{\dot{\theta}}{U}\bigg]
\end{equation}

\noindent Where the position of the rotation axis is identified with $x_f$, and the $x$ vector is positive in the flow direction. As, in our case, the rotation axis is placed at the quarter chord, a positive acceleration and velocity of the pitch degree of freedom will increase the coefficient of lift, thanks to the added mass effect and the dynamically induced angle of attack.

\nomenclature{$C_l$}{Lift coefficient}
\nomenclature{$c$}{Chord length}
\nomenclature{$C(k)$}{Theodorsen function}

\noindent In the figure, it can be seen a good match between the incompressible theory and the CFD computation. The same phase is predicted and only a small difference in the amplitude can be seen. This difference in amplitude is relatively common when comparing a prediction made with the thin foil assumption and a computation that takes into account the entire thickness or real experiments \cite{mccroskey_critical_1987}. This difference can also be seen in the previous results, where, in Table \ref{tab:naca_static}, the plunge displacement coming from SU2 is larger than the one predicted with the coefficients coming from X-FOIL and from the theory.

\subsubsection{Dynamic aeroelasticity} 

For this final example, two possible aerodynamic models have been considered. The first is based on the RANS equations used before. The other uses Euler equations for the aerodynamics. This has been done to completely verify the code for all the possible uses.

\noindent The Mach number will be gradually increased, with a fixed Re number of 4 millions and a fixed temperature of \SI{273.15}{K}, until the instability point of the aeroelastic system is reached. The classical pitch-plunge flutter will then be visible.

\noindent It must be noted that, to keep the Re number fixed in the RANS calculation, so that the same flow of energy from the momentum to the energy equations is present in all the calculations, the density of the flow must vary for the different Mach numbers.

\nomenclature{$\mathrm{Re}$}{Reynolds number}
\nomenclature{$\mathrm{Ma}$}{Mach number}

The equations presented at the beginning of this section, governing the airfoil dynamics, are usually adimensionalised to obtain results independent from the free-stream density of the flow.

\noindent Indeed, we can define the following parameters:

\begin{equation}
\chi=\frac{S_m}{mb},\ r_{\alpha}^2=\frac{I_f}{mb^2},\ \bar{\omega}=\frac{\omega_h}{\omega_{\alpha}},\ \mu_m=\frac{m}{\pi \rho_{\infty} b^2}
\end{equation}

\noindent Where $b$ is the semi chord of the airfoil, $\omega_h = \sqrt{\frac{K_h}{m}}$, $\omega_{\alpha} = \sqrt{\frac{K_{\alpha}}{I_f}}$. If we fix them, the structure will behave always the same regardless of $\rho_{\infty}$.

\noindent The selected parameters for this example are taken from~\cite{sanchez_assessment_2016}: $\chi=0.25$, $r_{\alpha}=0.5$, $\omega_{\alpha} = \SI{45}{rad/s}$, $\bar{\omega}=0.3185$ and $\mu_m=100$.

\nomenclature{$\chi$}{Normalized static unbalance}
\nomenclature{$b$}{Half-chord length}
\nomenclature{$r_{\alpha}^2$}{Normalized inertia}
\nomenclature{$\bar{\omega}$}{Uncoupled natural pulsation ratio}
\nomenclature{$\omega_h$}{Uncoupled plunging natural pulsation}
\nomenclature{$\omega_{\alpha}$}{Uncoupled pitching natural pulsation}
\nomenclature{$\mu_m$}{Mass ratio}

On the other hand, as the Re number does not appear in the Euler equations, in that case we can directly fix both the temperature and pressure, thus the density.

The different values of Mach numbers and density values, for RANS simulations, are reported in Table \ref{tab:rho}.

\begin{table}[ht]
\caption{Density values used for the RANS simulations of the NACA 0012 dynamic FSI.}
\label{tab:rho}
\centering
\begin{tabular}{@{}lc}
\hline
Mach			    & Density ($kg/m^3$)    \cr
\hline
0.1 				&	2.072				\cr
0.2					&	1.036				\cr
0.3					&	0.691				\cr
0.357				&	0.580				\cr
0.364				&	0.569				\cr
\hline
\end{tabular}
\end{table}

No structural damping is included and a time step of 1 millisecond is used. At each time step, the inner FSI convergence is based on a maximum norm of the structural displacement vector of 1 micrometer.

\begin{figure}[htp]
\centering
\includegraphics[width=0.40\textwidth]{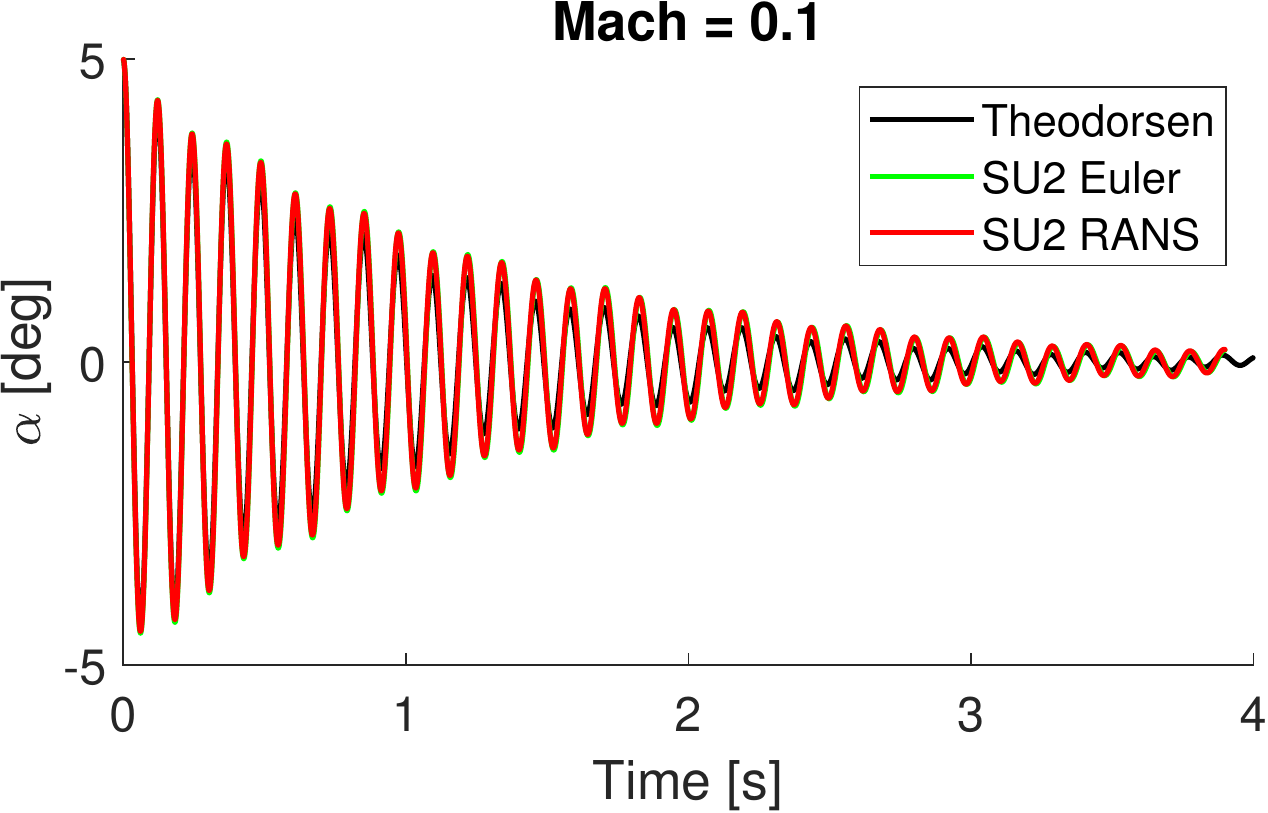} \quad
\includegraphics[width=0.40\textwidth]{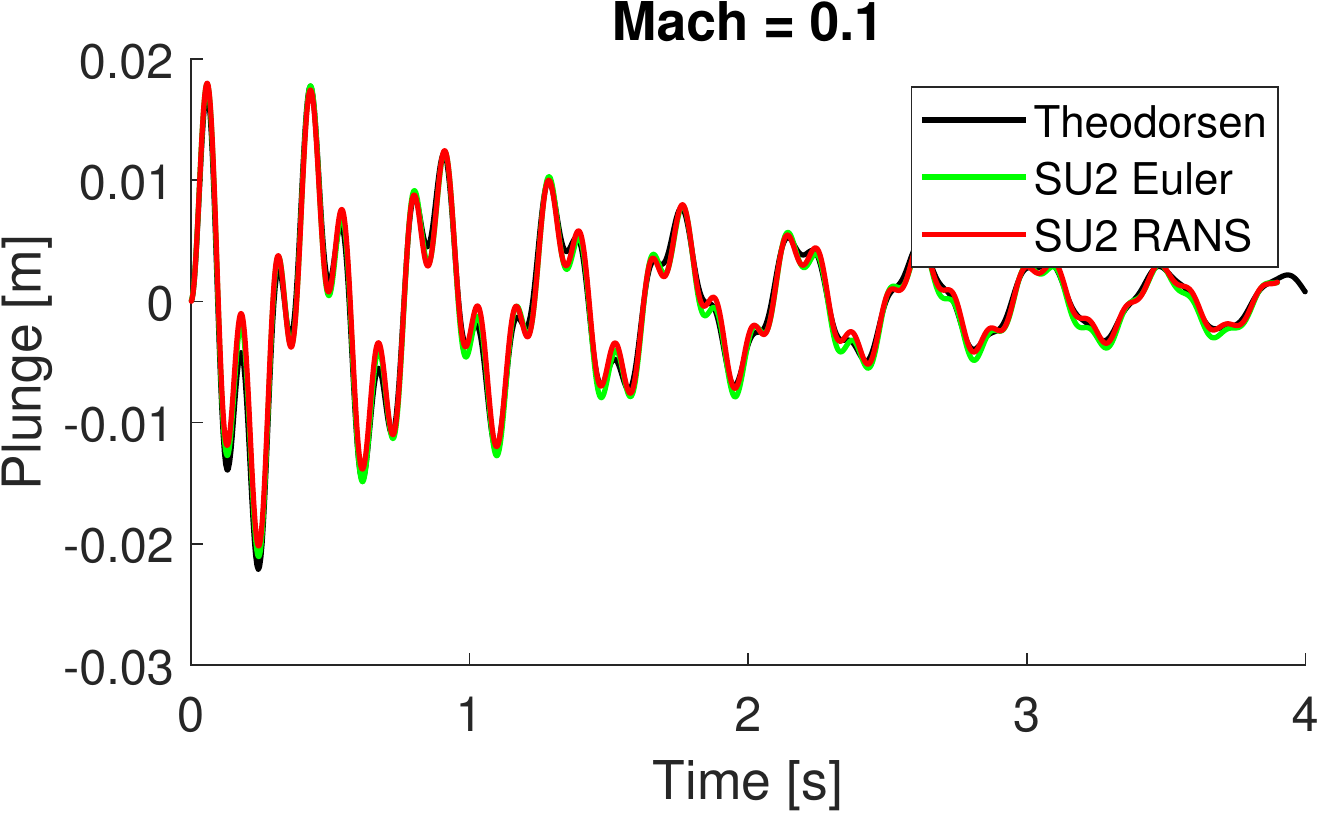} \vspace{0.2cm}
\includegraphics[width=0.40\textwidth]{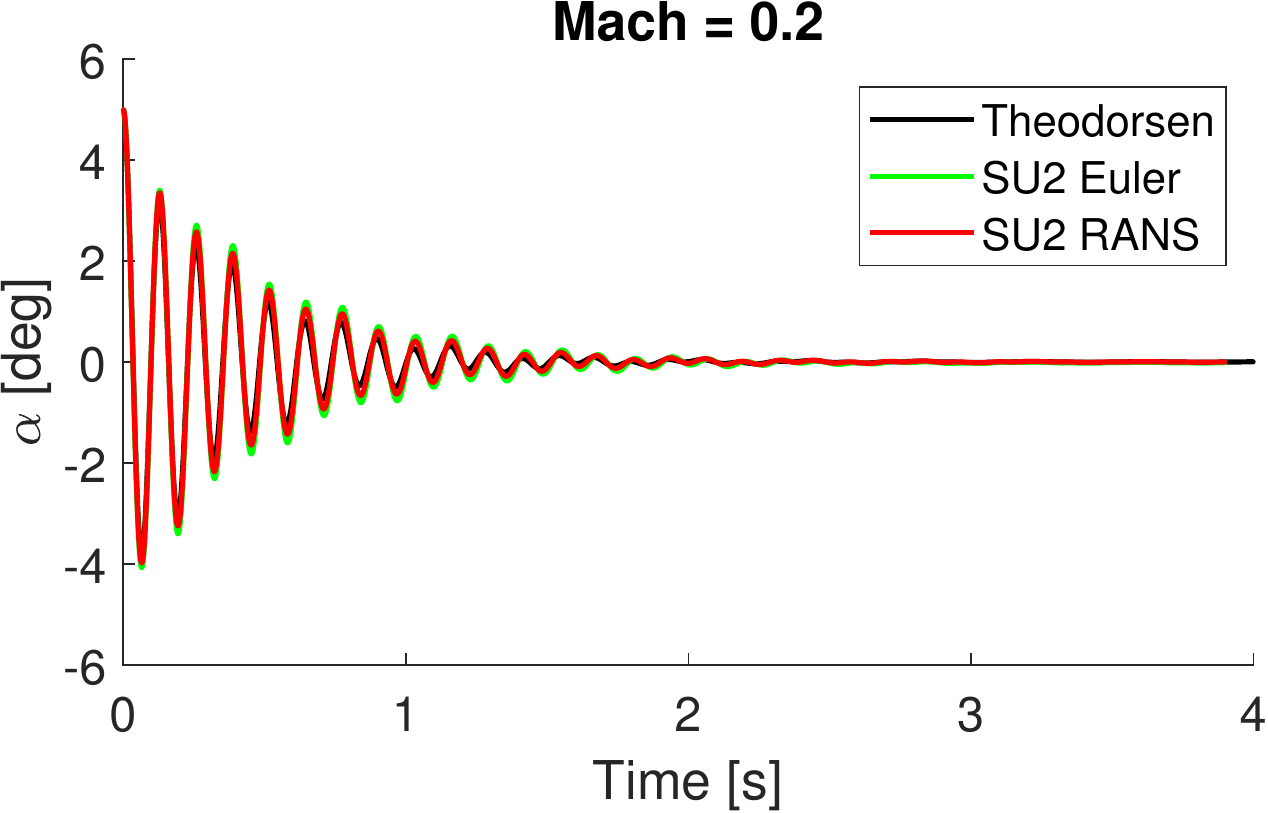} \quad
\includegraphics[width=0.40\textwidth]{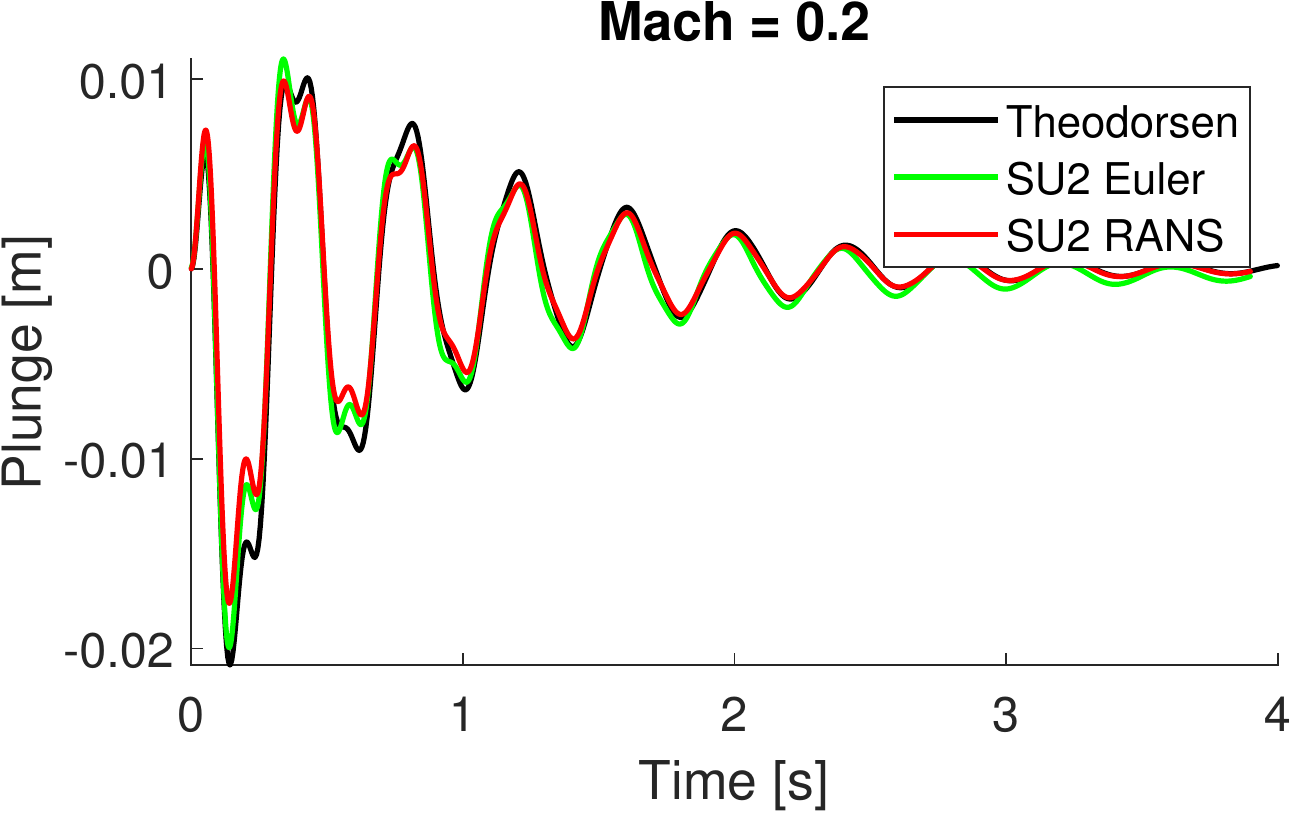} \vspace{0.2cm}
\includegraphics[width=0.40\textwidth]{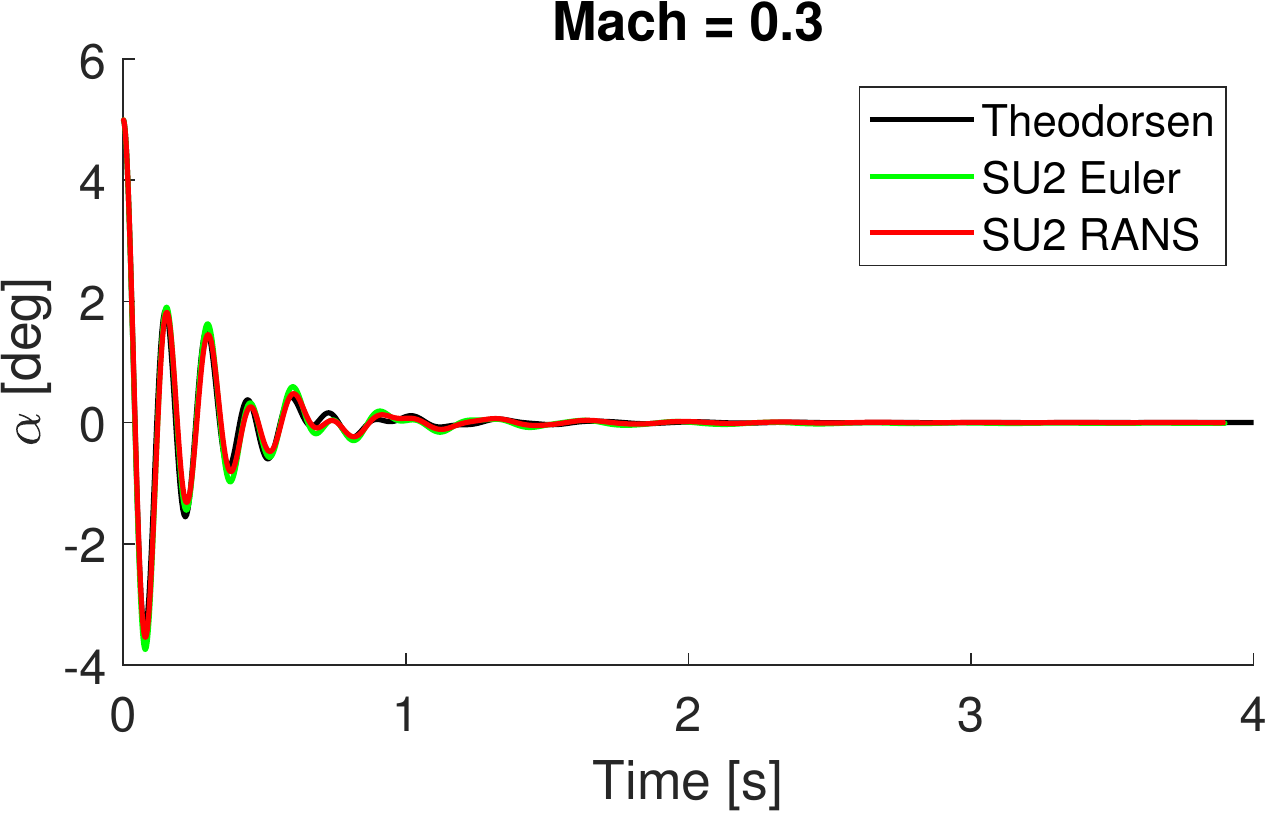} \quad
\includegraphics[width=0.40\textwidth]{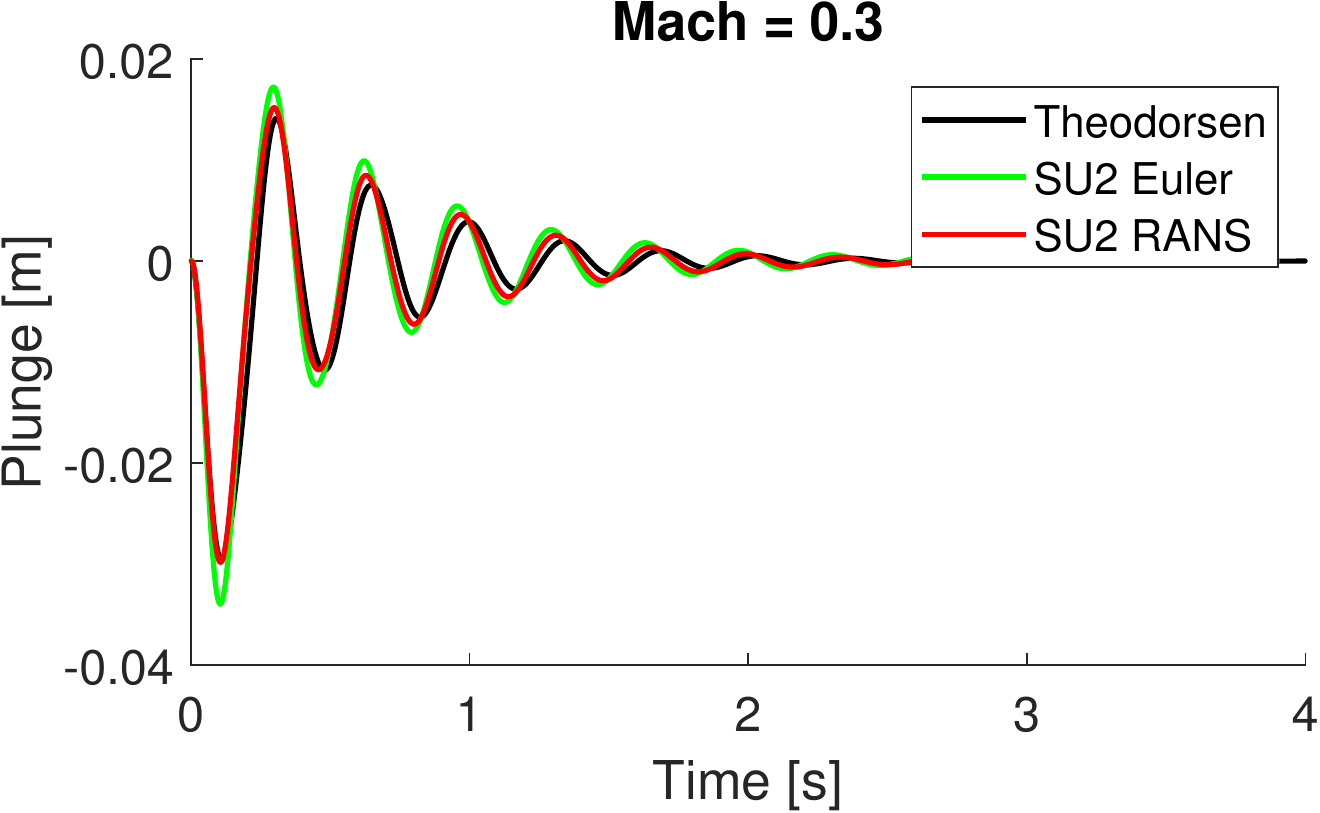} \vspace{0.2cm}
\includegraphics[width=0.40\textwidth]{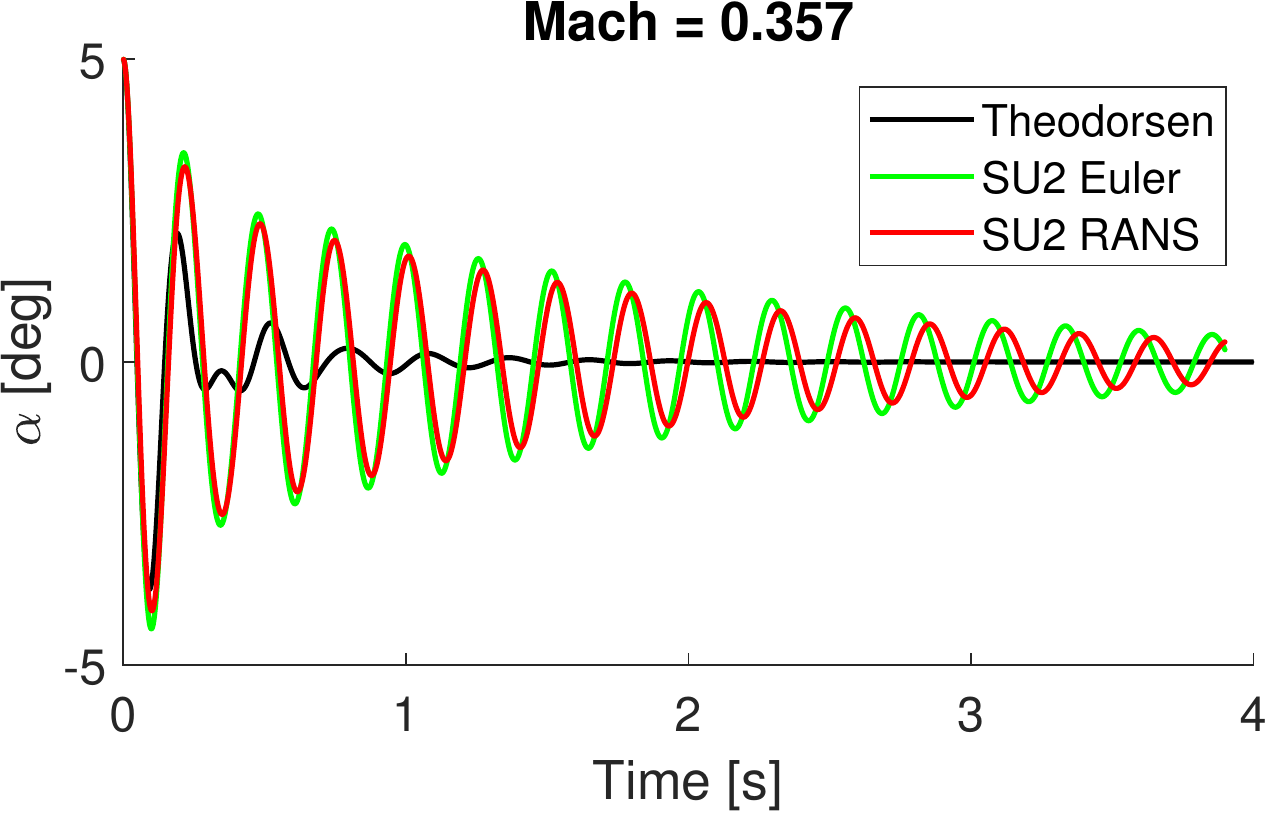} \quad
\includegraphics[width=0.40\textwidth]{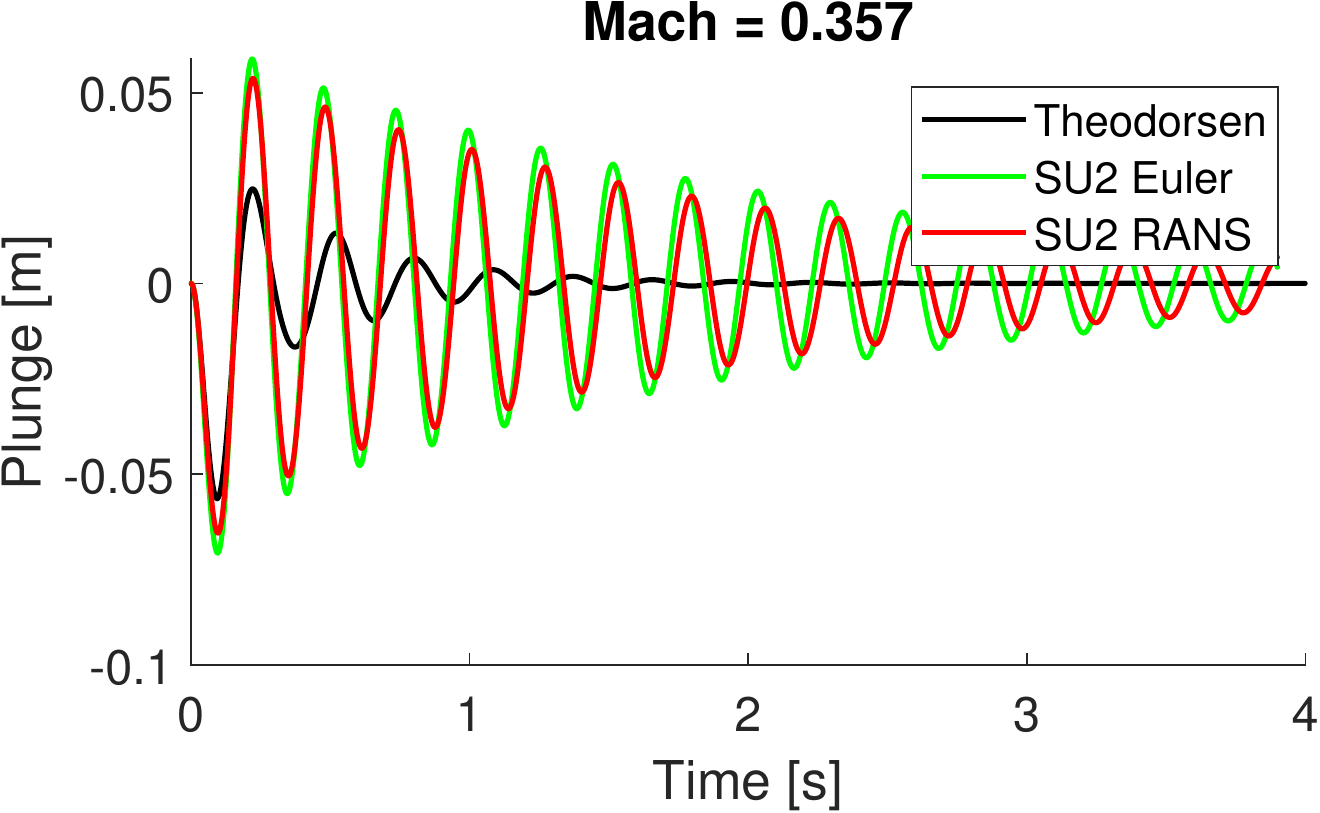} \vspace{0.2cm}
\includegraphics[width=0.40\textwidth]{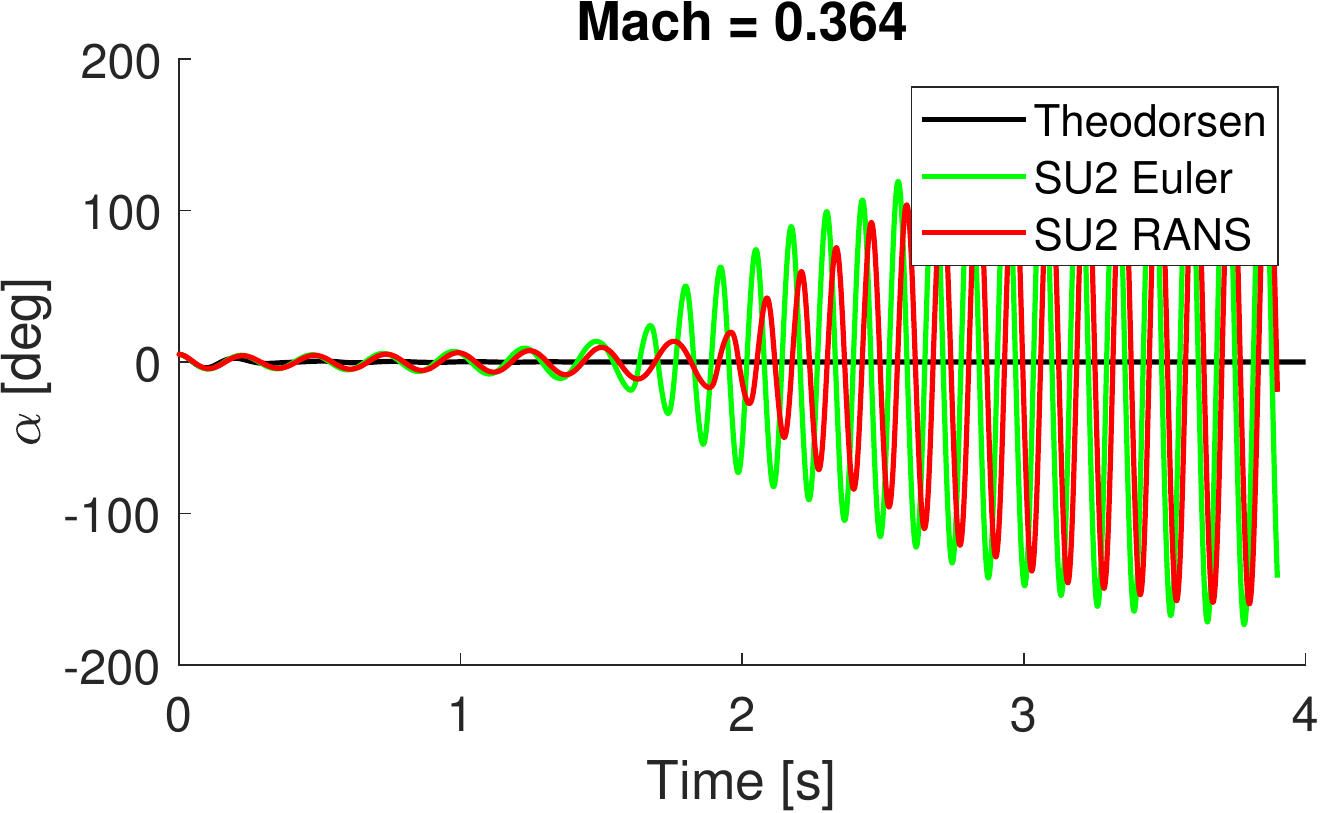} \quad
\includegraphics[width=0.40\textwidth]{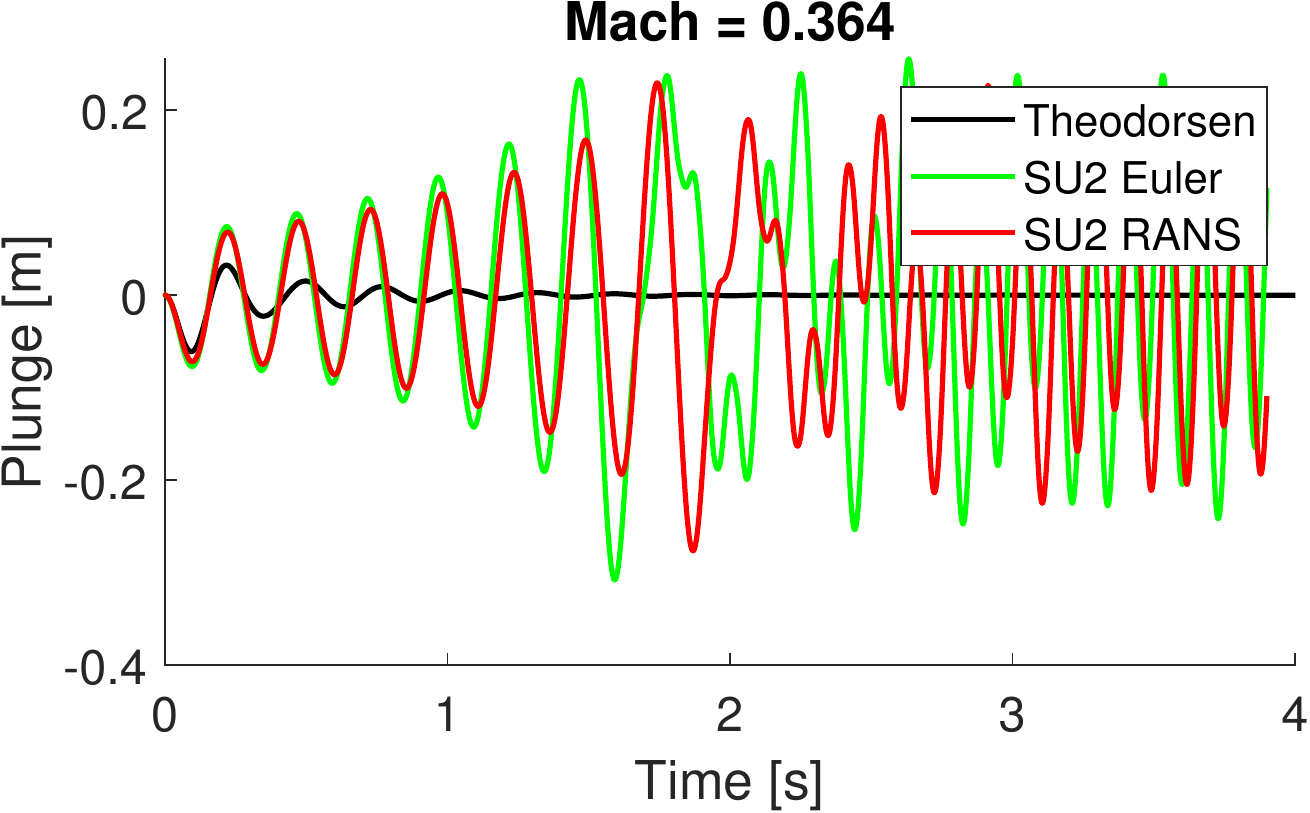}
\caption{Time domain responses of the NACA 0012 airfoil to an initial condition in pitch of $5^{\circ}$, for different Mach numbers.}
\label{fig:naca_flutter_time}
\end{figure}

\begin{figure}[ht]
\centering
\includegraphics[width=0.7\textwidth]{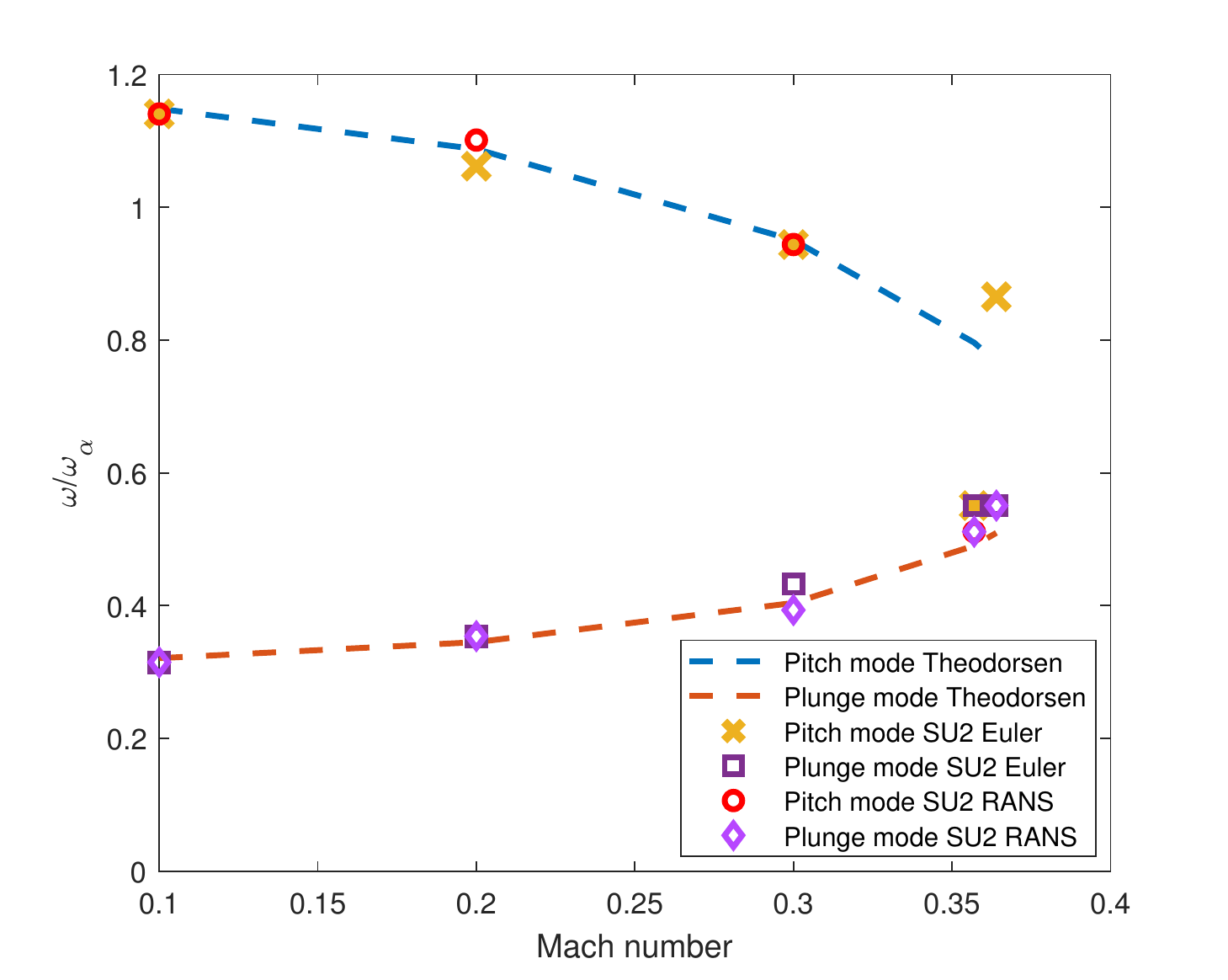}
\caption{Modal frequencies as a function of the Mach number for the NACA 0012 airfoil.}
\label{fig:naca_flutter_freq}
\end{figure}

\noindent The comparison of the time histories is shown in Fig. \ref{fig:naca_flutter_time}. The results in terms of frequencies are reported in Fig. \ref{fig:naca_flutter_freq}.

It can be seen how the frequency merging is well captured; after the flutter point the two frequencies are coincident and nonlinear effects are present. Thus, comparing the Theodorsen theory and SU2 is not fully meaningful. 

\noindent The time domain histories match nicely until the flutter point is approached. Here, both because the Mach is higher and nonlinear effects start to appear, and because the thick CFD model predicts flutter sooner, the solutions start to diverge. In general, for the first three Mach numbers, the match is good. As often happens, the Euler equations overestimate the lift peaks, while including viscosity the results are closer to the incompressible theory. The fourth Mach number is clearly at the edge of flutter, with pitch and plunge moving at a very similar frequency, and decaying slowly. It can also be seen that the Euler simulation seems to tend more to the instability. The last Mach number is past the flutter point. Very large oscillations are present, but they are not physical. Indeed, we should recall that the structural model is based on a linear FEM, which fails when large displacements or rotations are involved. Confirming our predictions, Euler equations predict a faster increase in the LCO amplitude. The comparison with the thin airfoil, small displacement, theory is not meaningful at this point.

\subsection{Benchmark SuperCritical Wing}

In this section, a more complex case will be tackled. From the structural point of view, no much complicacy is added. However, from the aerodynamic point of view, the flow is significantly more nonlinear, in the transonic range, exhibiting high unsteadiness and continuous shocks formation. The case is commonly known as the BSCW \cite{heeg_investigating_2017}. The wing is a semirigid, semispan, rectangular wing based on the SC(2)0414 supersonic airfoil, with two degrees of freedom in pitch and plunge. The chord is $\SI{0.4064}{m}$ ($16$ in.) and the span is $\SI{0.8128}{m}$ ($32$ in.).

\begin{figure}[ht]
\centering
\includegraphics[width=0.5\textwidth]{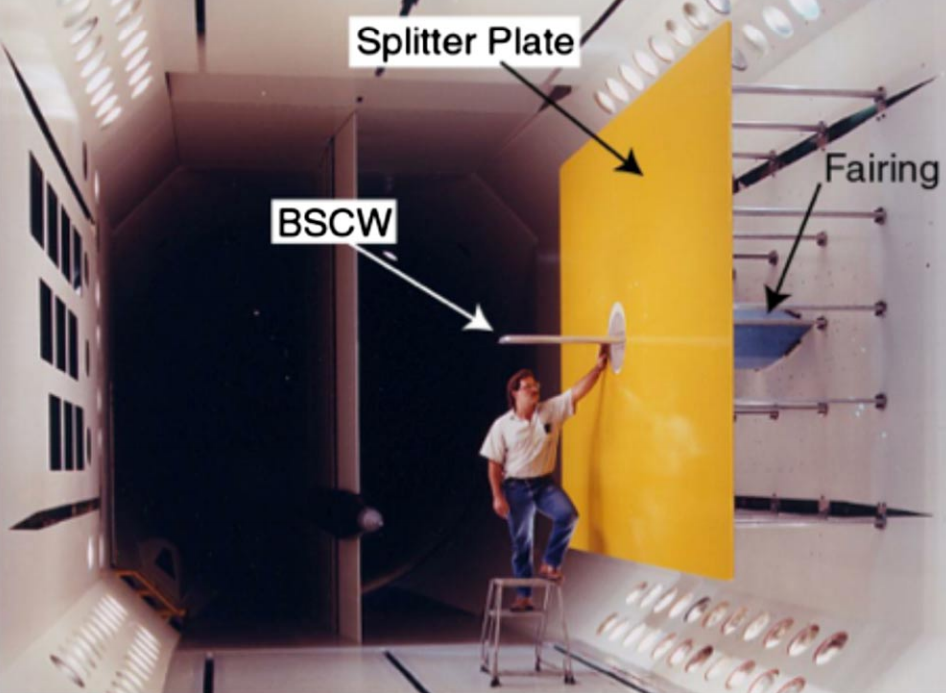}
\caption{Physical apparatus used in the experiments for the BSCW, taken from \cite{heeg_plans_2015}.}
\label{fig:bscw_mounted}
\end{figure}

The analyses considered in this section reproduce two of the reference test cases used for the Aeroelastic Prediction Workshop~\cite{heeg_overview_2016}, whose results can be compared with wind-tunnel experiments of the BSCW model that were conducted in the NASA Langley Transonic Dynamics Tunnel. A picture of the wing mounted on the suspending apparatus is reported in Fig. \ref{fig:bscw_mounted}, taken from \cite{heeg_plans_2015}.

\noindent The first case here reported is a forced excitation analysis, in which an oscillating turntable varies the pitch of the wing about an axis at the $30\%$ of the chord~\cite{heeg_experimental_2013}.
The second case is the flutter analysis of the pitching and plunging wing~\cite{dansberry_experimental_1993}; unlike the previous case, the pitch motion is about the center chord.
Although the BSCW was semirigid~\cite{dansberry_experimental_1993}, the simulations are performed on a rigid wing, with a single dof in the forced case and two dofs in the flutter case.

From the structural modeling point of view, the system is made of one master node only, with two degrees of freedom. These correspond to the pitching and the plunging of the wing. Several slave nodes are connected rigidly to the master node, and are positioned so to represent the thickness and the span of the three--dimensional wing, as previously done for the test case of the NACA airfoil. In order to retain stability, they cannot all lie in one single plane. Lumped masses, positioned at the slave nodes, are used to represent the mass distribution of the model, and two spring elements are placed on the free degrees of freedom of the master node to represent the elastic properties.

The aerodynamic model is significantly more complex, due to the physical phenomena to be represented. RANS equations will be used, with a standard Spallart-Almaras turbulence model, as large regions of separation are not expected to occur, up to the instability point. The physical thermal conductivity is modeled with a constant Prandtl number of 0.755, while the turbulent conductivity is not modeled.

\begin{figure}[ht]
\centering
\includegraphics[width=0.26\textwidth]{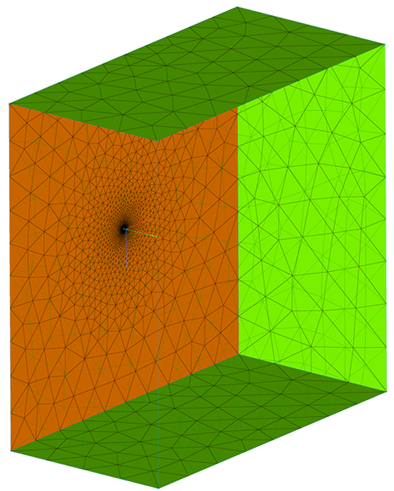}\quad
\includegraphics[width=0.7\textwidth]{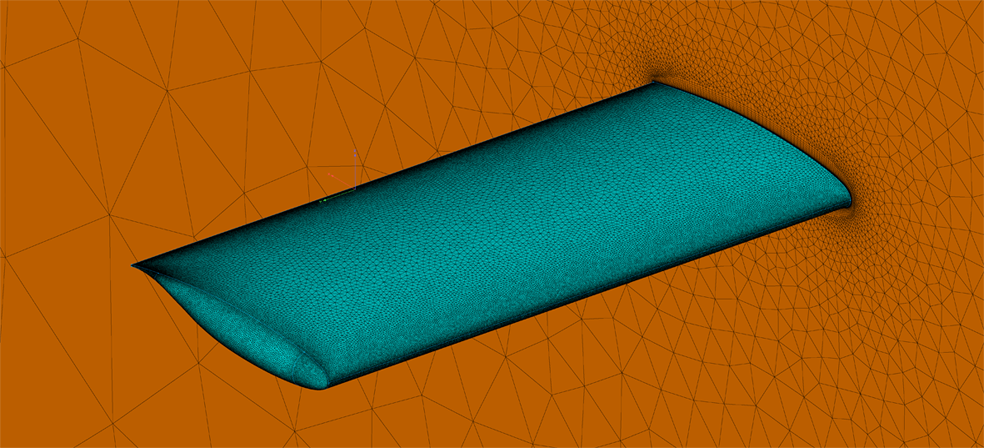}
\caption{Aerodynamic mesh for the RANS simulation of the BSCW. Euler boundary condition in orange, far-field boundary conditions in green, no-slip boundary condition in blue.}
\label{fig:bscw_cfd_mesh}
\end{figure}

\noindent The aerodynamic boundary conditions do not take into account the presence of the wind tunnel itself, as this is slotted and there should not be any shock reflection. Only the splitter plane is modeled, using a no penetration boundary condition. The other external boundaries are placed far from the wing, at approximately 100 chords from it, and are modeled with a far-field Riemann boundary condition. At the wing, no-slip is imposed, and cells close to the surface are placed at $y^+\approx 1$. The mesh is unstructured, with 3 million nodes and prison layers close to the surface, and it is reported in Fig. \ref{fig:bscw_cfd_mesh} together with the identification of the boundary conditions.

As the computation is now significantly more expensive, an adaptive strategy for the CFL number is used. The minimum is set to 1, while the maximum is set to 100. Due to the possibility of having a relatively large local pseudo time, the Bi-Conjugate Gradient method is used for the linear solver. A physical time step size of \SI{1}{ms} is used.

\subsubsection{Forced motion aerodynamics}

In this forced response case, the wing is oscillated in pitch, with an amplitude of 1 degree at $\SI{10}{Hz}$.
\noindent The analysis is performed in R-134a gas at Mach 0.7, Re number equal to 4.5 millions and a 3 degrees mean angle of attack.

\begin{figure}[ht]
\centering
\includegraphics[width=0.65\textwidth]{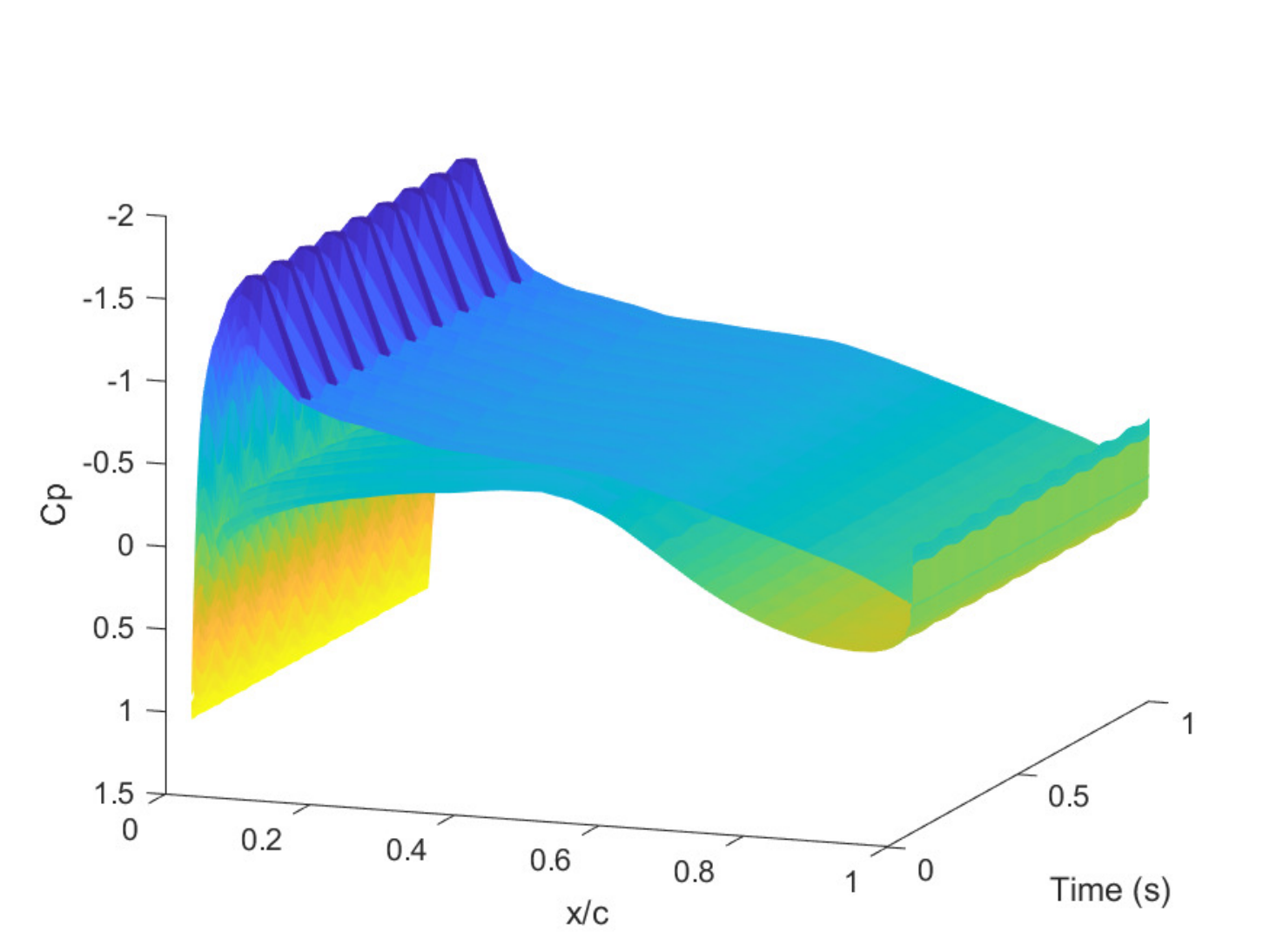}
\caption{Coefficient of pressure distribution at the $60\%$ span station of the BSCW. Imposed sinusoidal pitching motion with $3^{\circ}$ mean angle of attack, $1^{\circ}$ amplitude, and $\SI{10}{Hz}$ frequency.}
\label{fig:bscw_forced_cp}
\end{figure}

\noindent The available experimental data is the unsteady pressure distribution at the $60\%$ span station. Therefore, the results reported here are limited to that station.
Fig.~\ref{fig:bscw_forced_cp} depicts the variation in time of the pressure coefficient at the $60\%$ span station. It can be noted the formation of a shock wave at the suction side, during the pitching motion. Indeed, it can clearly be seen in Fig. \ref{fig:bscw_forced_cp} how, when the angle of attack approaches the value of $4^{\circ}$, a shock forms at around 10 percent of the chord.

\begin{figure}[ht]
\centering
\includegraphics[width=0.9\textwidth]{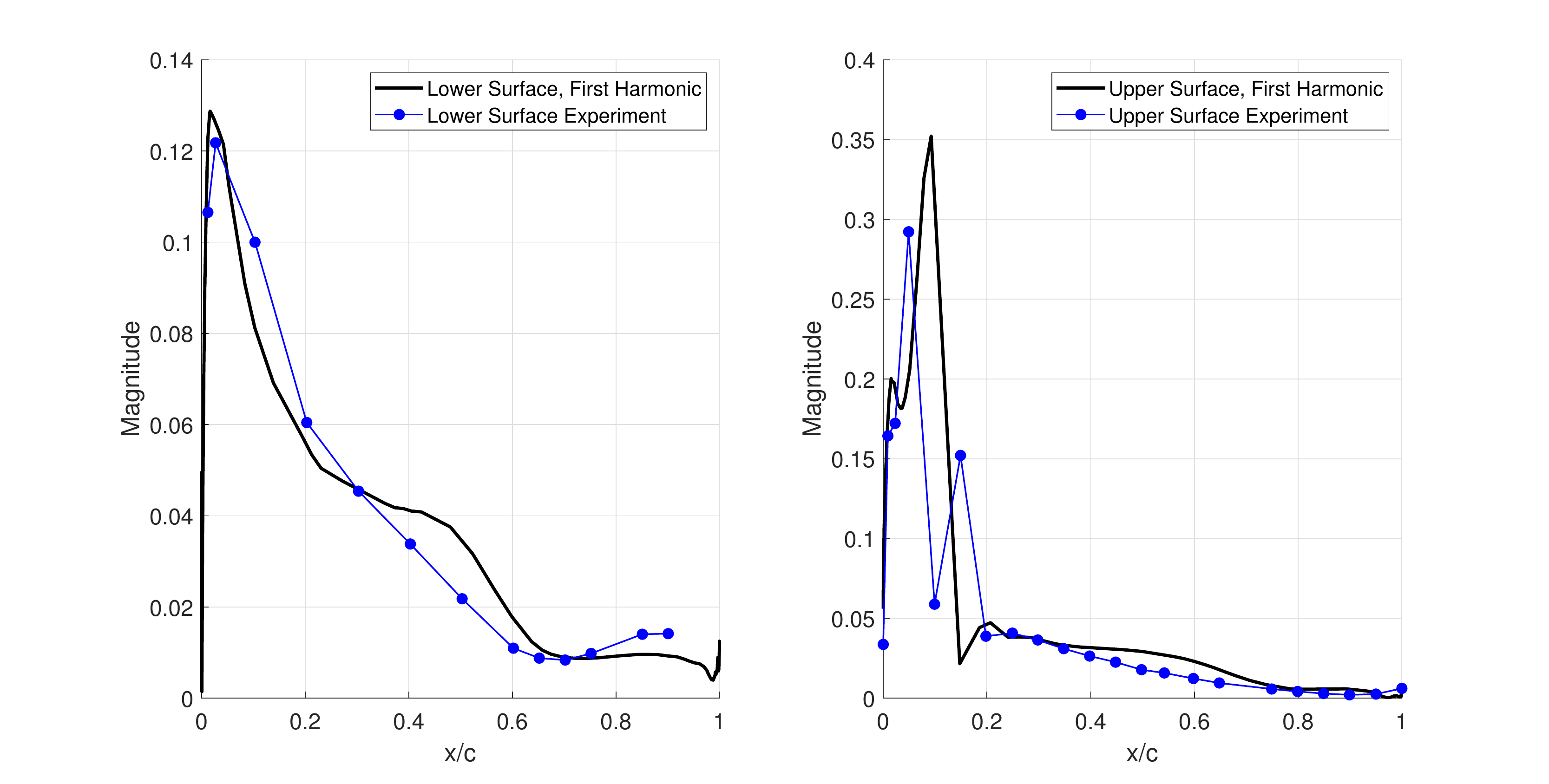}
\caption{Transfer function magnitude between the coefficient of pressure of the BSCW at the $60\%$ span station and the pitch rotation. Imposed sinusoidal pitching motion with $3^{\circ}$ mean angle of attack, $1^{\circ}$ amplitude, and $\SI{10}{Hz}$ frequency.}
\label{fig:bscw_forced_tf_ampli}
\end{figure}

\begin{figure}[ht]
\centering
\includegraphics[width=0.9\textwidth]{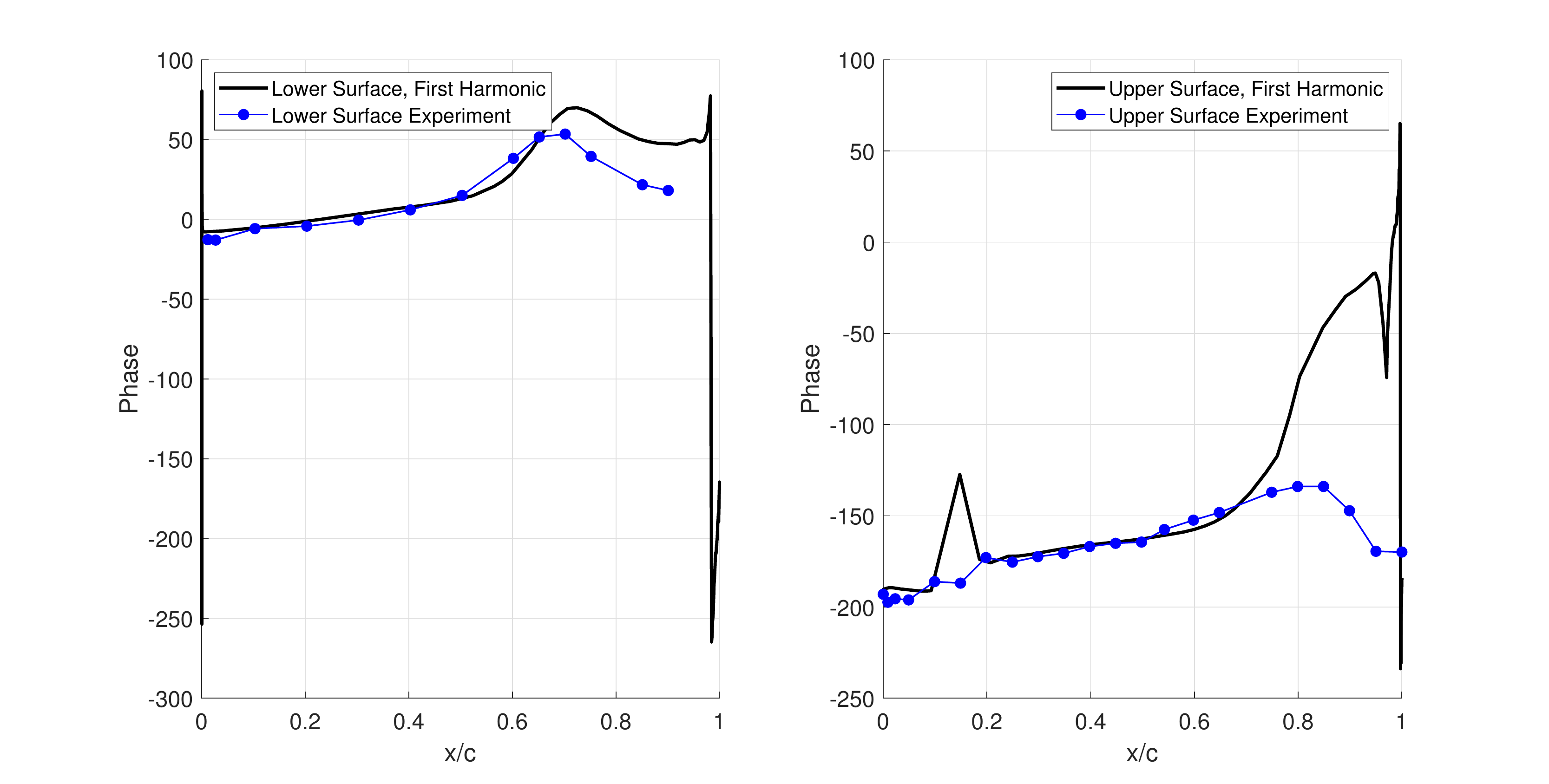}
\caption{Transfer function phase between the coefficient of pressure of the BSCW at the $60\%$ span station and the pitch rotation. Imposed sinusoidal pitching motion with $3^{\circ}$ mean angle of attack, $1^{\circ}$ amplitude, and $\SI{10}{Hz}$ frequency.}
\label{fig:bscw_forced_tf_phase}
\end{figure}

\noindent The transfer function between the pressure coefficient and the pitch angle is also computed. It is then compared with that obtained from the experimental data. Fig.~\ref{fig:bscw_forced_tf_ampli} represents the amplitude of the transfer function for both the upper and lower surfaces. The phase is reported in Fig.~\ref{fig:bscw_forced_tf_phase}.

The obtained results are really close to those in the available literature. An excellent review paper, prepared after the second aeroelastic prediction workshop \cite{heeg_overview_2016}, compares results obtained by all the participants at the workshop and provides the required information to understand the present example. In all the simulations, it can be seen how most of the amplitude response is at the leading edge, and this is correctly predicted by the numerical methods, including the present one. Two peaks are visible on the upper side. The first, is due to the acceleration of the flow as the wing pitches up, and the deceleration when the wing pitches down. This is due to the geometry of the airfoil only. The second, higher, is due to the formation of the shock wave. The flow is attached, as it can be seen in the phase plot. Indeed, on the upper surface, with a positive pitch up movement, we have a decrease in pressure coefficient (phase is $-180^{\circ}$), while on the lower side the opposite is true. A peak in the phase can be observed in the upper surface, after the shock. This may be due to a small separation bubble formed by the shock itself that creates delay in the response of the pressure coefficient to pitching.

\noindent On the lower side, only one peak in the magnitude plot is visible, as there is no shock wave formation.

Discrepancies are found between simulations and experimental values close to the shock wave. One sensor on the upper surface is not matching simulation values. The same discrepancies have been found by all the other participants to the workshop, and have been attributed to a fault sensor. 

\noindent In the phase plot, differences arise also close to the trailing edge due to the transfer function going to zero. Under these conditions, the ratio between imaginary and real part of the transfer function is ill conditioned and the computation prone to error.

As stated before, the results well match others in the literature and confirm the applicability of the developed framework for complex, 3D, aerodynamically nonlinear cases.

\subsubsection{Dynamic aeroelasticity}

The final test for the current framework was the search of the flutter point for the BSCW wing. This has also been called the case 2, in the second aeroelastic prediction workshop. Experimental results are available and show a flutter dynamic pressure of 8082 Pa at Mach number 0.74, Re number equal to 4.4 millions, and no angle of attack.

\noindent The model is characterised by a plunge mode with a frequency of $\SI{3.3}{Hz}$ and a pitch mode with a frequency of $\SI{5.2}{Hz}$.

\begin{figure}[ht]
\centering
\includegraphics[width=0.47\textwidth]{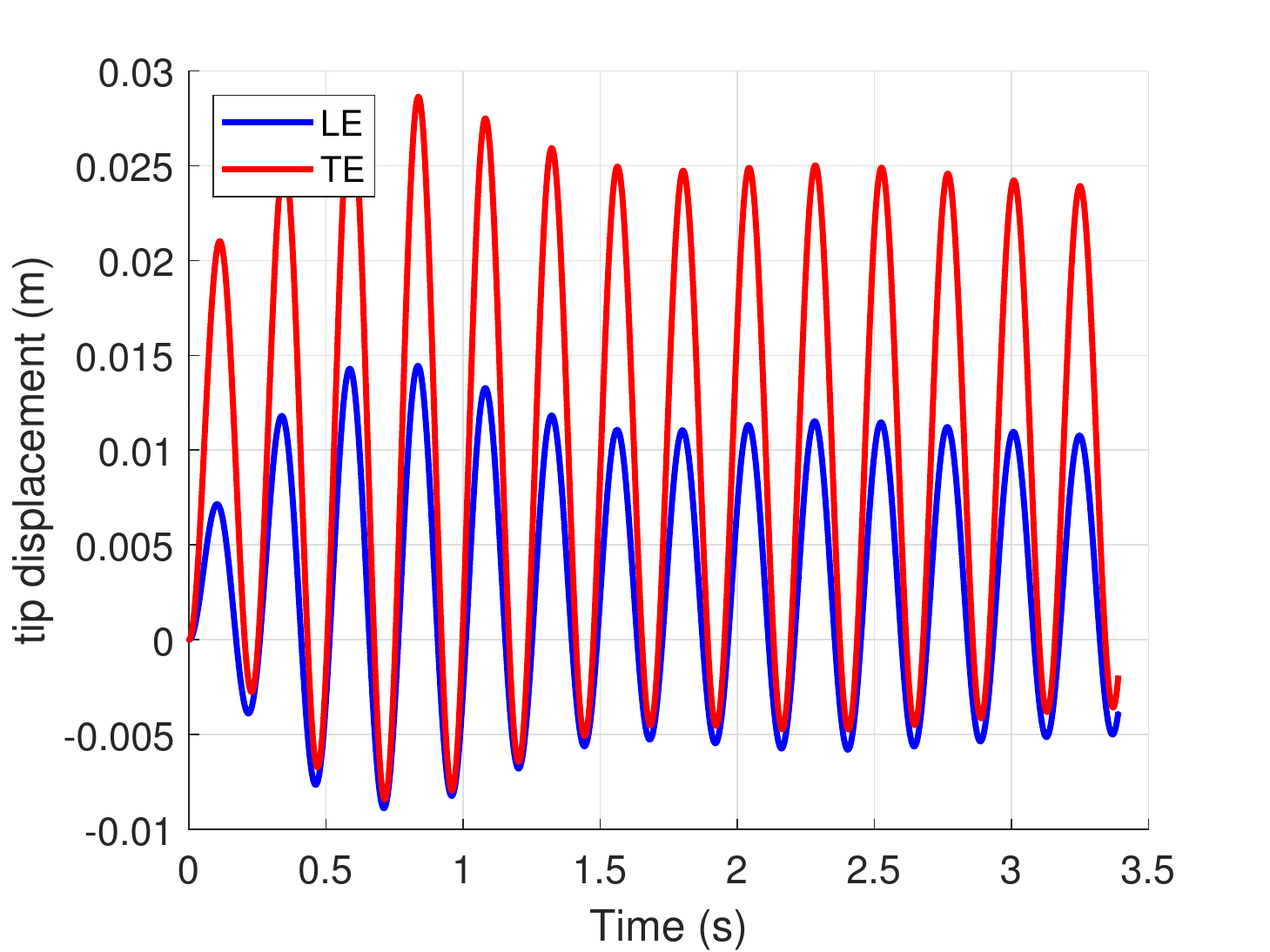}\quad\quad
\includegraphics[width=0.47\textwidth]{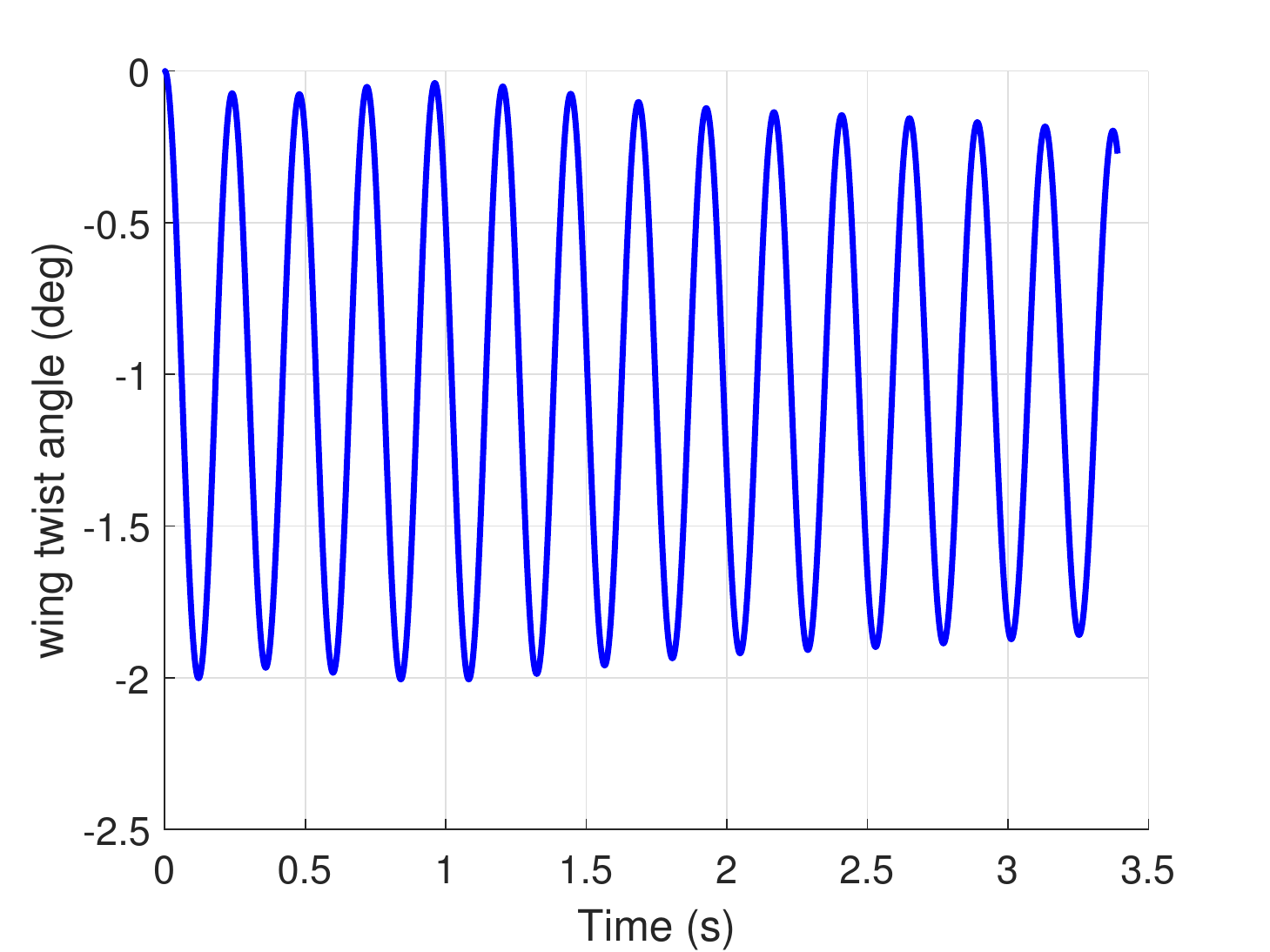}
\caption{Time histories of the wing displacements and wing twist for the experimental flutter point.}
\label{fig:BSCW_beforeflutter}
\end{figure}

First, a coupled simulation is run using the experimental values for the flutter point. Results, in terms of displacements of the wing, are reported in Fig. \ref{fig:BSCW_beforeflutter}. It can be seen that the system is only lightly damped, confirming the proximity of the flutter point. However, it is stable.

\begin{figure}[ht]
\centering
\includegraphics[width=0.45\textwidth]{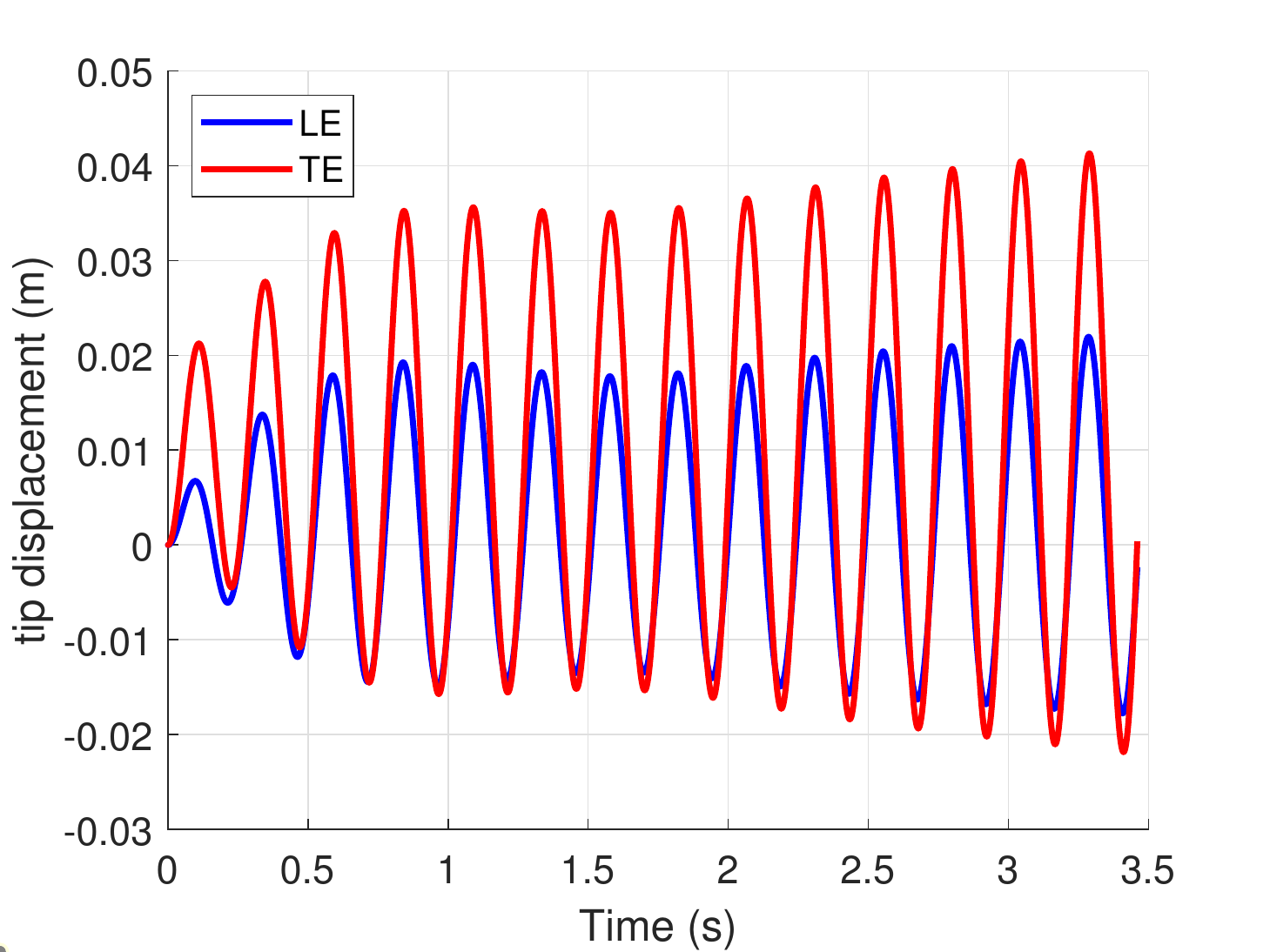}\quad\quad
\includegraphics[width=0.45\textwidth]{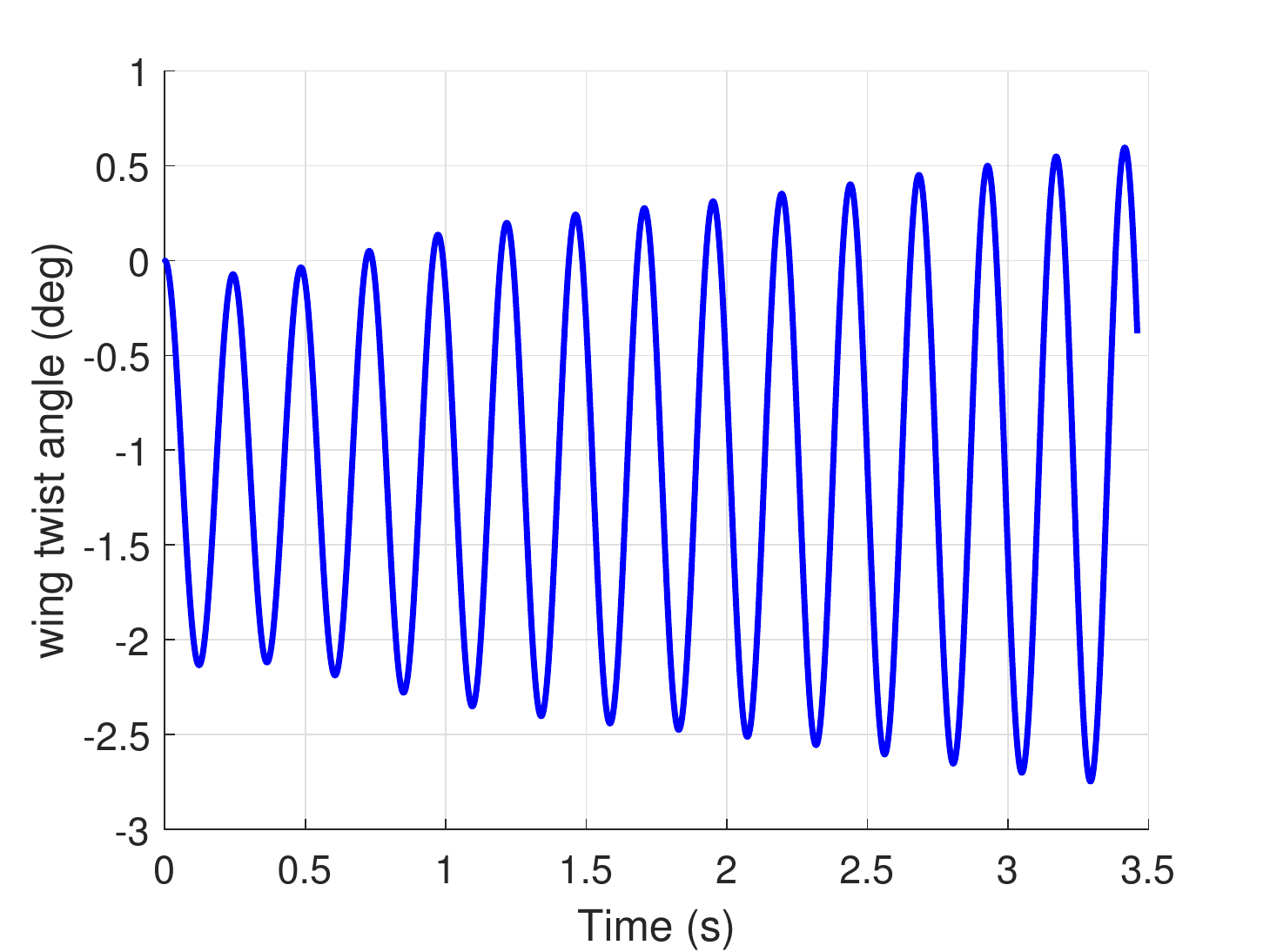}
\caption{Time histories of the wing displacements and wing twist for the numerical flutter point.}
\label{fig:BSCW_afterflutter}
\end{figure}

In order to test whether the code could actually predict an unstable behaviour for the BSCW, the dynamic pressure was increased by 5\%. The simulation is then run again, with the same parameters except for the new dynamic pressure, and the time histories are reported in Fig. \ref{fig:BSCW_afterflutter}. It can be clearly seen an unstable behaviour with the displacements growing in amplitude, especially for the wing twist. This confirms that the numerical flutter point is in good agreement with the experimental one, and it also confirms the capabilities of the developed code.

\subsection{Transonic wind tunnel test simulation of a flexible demonstrator}

The Clean Sky 2 European research project GUDGET aims at the design and manufacturing of an innovative experimental set--up for the investigation of gust loads in transonic flow conditions. The experimental set--up is composed of a gust generator and an aeroelastic half--model to be installed in the transonic ONERA S3Ch wind tunnel facility. One of the main goals of the wind tunnel tests is to study non--linear effects induced by high amplitude gust loads on the aeroelastic behavior of the model. Since the structural behavior is expected to be linear, the numerical verifications should be performed considering the interaction between a linear finite element model and an unsteady aerodynamic model able to capture non--linearities effects. Thus, it is the perfect application of the present code.

\noindent The wind tunnel is a transonic facility with a 0.8 m x 0.8 m square test section of 2.2 m length. It covers a Mach number range from 0.3 to 1.2 and operates at atmospheric stagnation pressure and temperature. The aeroelastic half--model consists of a wall--mounted fuselage connected to a swept wing. The aerodynamic wing shape is based on the supercritical OAT15A airfoil \cite{rodde_selection_1994}. The swept angle is equal to 30 deg, while the wing twist ensures a shock wave parallel to the leading edge, as well as a constant pressure along the span in cruise conditions, without separation at the wing root. The chord varies from 204.55 at the root to 170.45 mm at the tip, over a span of 600 mm which limits wall effects while maximizing the local thickness. The aerodynamic reference condition is characterised by a freestream Mach number of 0.82.

\noindent The aeroelastic model is equipped with a classical aileron able to dynamically move with a bandwidth which is compatible with the frequency content of the gust generator. The hinge axis is located at 75\% of the local chord, between 65\% and 85\% of the wing span \cite{molton_control_2013}.

\noindent The model can be connected to the wind tunnel in three ways. First, with a rigid connection, used to impose a precise angle of attack. Second, with an elastic connection designed so that the pitch mode of the demonstrator is at 20 Hz. Finally, with a motorised connection, able to force the pitch with varying frequency and amplitude \cite{lepage_complete_2015}.

A linear unsteady aerodynamic method cannot take into account the non--linearities at the considered reference Mach number. For this reason, the FSI approach based on CFD is adopted as higher fidelity method for flutter verification. It must be pointed out that the CFD computation is used to include the non--linearities effects of the aerodynamics, but the finite element model is still considered linear.
After flutter studies, the stability margin with respect to the position of the rotation axis is investigated, considering three different positions in wind direction.

\subsubsection{Structural modeling}\label{sssec:GUDGstructModel}

The wing of the complete wind tunnel model is made of a single piece, milled from solid, providing a common steel platform. This platform is directly connected to pitch shaft on one side of the fuselage. Multiple cover parts are connected at the lower surface of the wing where instrumentation and the actuated aileron is to be installed. The aileron linkage also consists of a complete single piece mechanism, equipped with compliant hinges, which excludes any free play, but adds stiffness to the actuation kinematics. All cover parts restore the structural stiffness of the wing and make it locally lighter. 

The finite element model is based on a continuous plate discretization able to reproduce the stiffness distribution of the actual wing, considering all local details described above. In this plate model, the camber distribution represents the mean surface of the wing, while the thickness distribution is used to assign thickness properties to each plate element by a mapping able to extract the information from the solid CAD model. This plate model is then tuned in terms of mass distribution matching the inertial properties obtained by the CAD model.
The fuselage is considered as rigid, and its inertial properties are introduced into the model by lumped masses placed in the center of gravity of each mechanical component included in the CAD model. The complete model is balanced by two additional masses placed in the front and the back of the pitch axis position. All these lumped masses are rigidly connected to the rotational axis together with additional nodes that are used for the interpolation of the fuselage displacements on the aerodynamic mesh.

\noindent The complete finite element model is validated using the vibration modes computed on a full--solid model consisting of about half million tetrahedral elements and not suitable to perform aeroelastic analyses coupled with CFD computations. Before using the plate model to verify the flutter behavior in the transonic reference condition, several parametric analyses were performed for different values of the stiffness to be assigned to the 2 aileron actuators and to the pitch mechanism: values close to what is expected from the model design are 3600 Nm/rad for the pitch mode and a total value of 31.8 Nm/rad for the aileron.

\subsubsection{Aeroelastic modeling}

The aeroelastic model consists of a modal base, reduced to the first six vibration modes computed on the finite element plate model described in section \ref{sssec:GUDGstructModel}, coupled with a CFD model prepared starting from the external surfaces of the CAD model.

\noindent The CFD mesh, obtained after a convergence study, is made of an unstructured block, refined at the wing surface to include 50 prison layers, required for the boundary layer resolution, the first of which is placed at $y^+\approx 1$. The mesh consists of 2 million nodes and represents the wind tunnel test section, which is a rectangular channel, and the half-plane model. In order to avoid spurious effects due to the boundary conditions, the test section is increased in length to 9 meters. No penetration boundary conditions were imposed on all the walls of the channel, no-slip boundary condition was imposed on the model surface, and the total thermodynamic quantities of pressure and temperature were imposed at the inlet. The outlet static pressure was then imposed so that in the middle of the channel the Mach number was equal to 0.82. Results of a steady simulation at the prescribed Mach number are reported on the left of Fig. \ref{fig:Cp_compare}. Here, the coefficient of pressure is plotted over the surface. It can be seen that, as already mentioned in the introduction, the wing has been designed to have a smooth variation of the shock location along the span. Further, the root section is free from shock thanks to appropriate twisting of the wing. Due to the separation present, SST was chosen as turbulence model.

\begin{figure}[ht]
\centering
\includegraphics[width=0.8\textwidth]{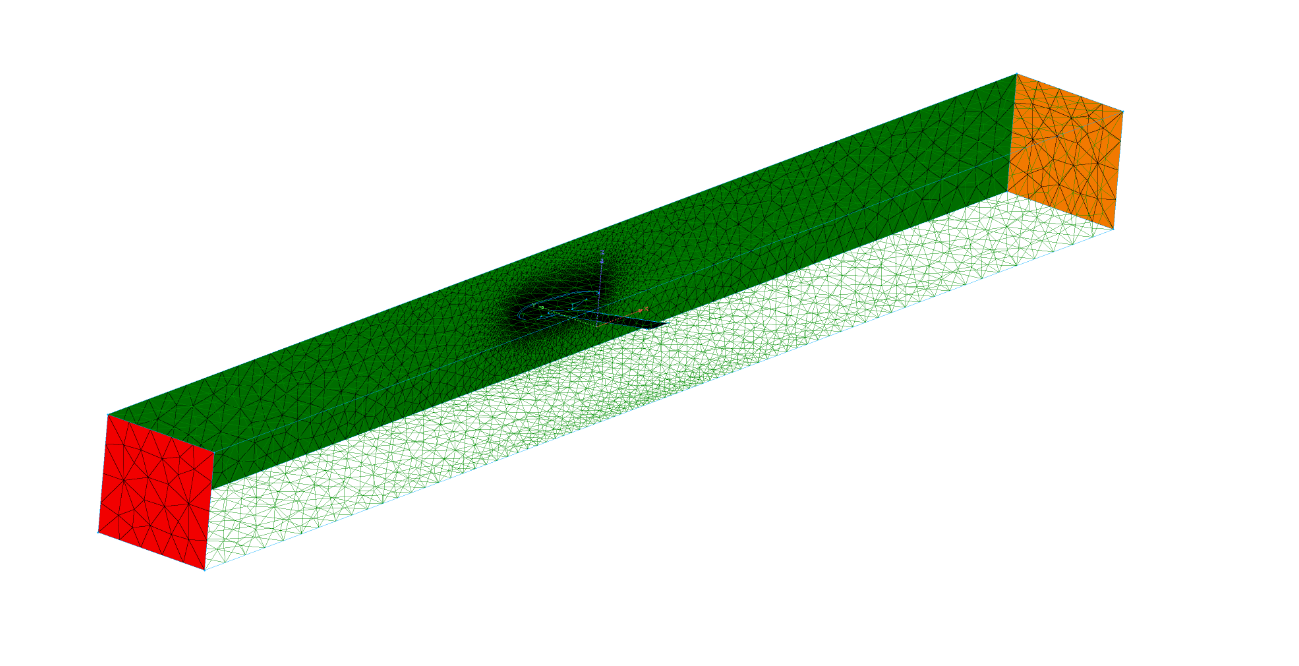}
\caption{Aerodynamic mesh for the GUDGET model in the wind tunnel. Red surface identifies the inlet, orange identifies the outlet, and green the no-penetration boundary conditions.}
\label{fig:mesh_gudget_far}
\end{figure}

\begin{figure}[ht]
\centering
\includegraphics[width=0.75\textwidth]{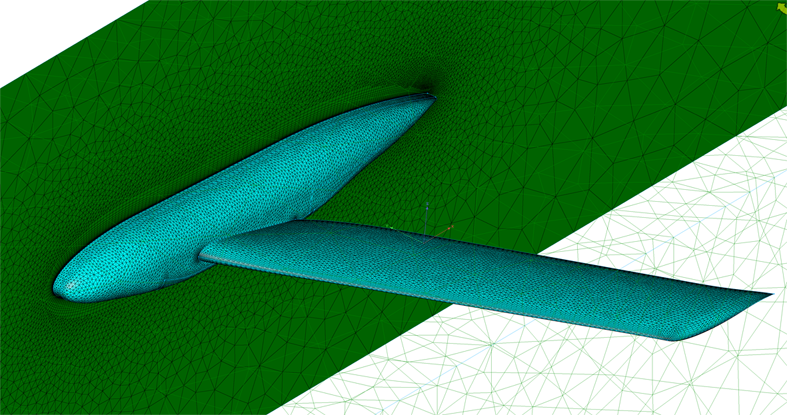}
\caption{Close-up view of the aerodynamic mesh for the GUDGET model in the wind tunnel. Light blue surfaces identify the no-slip boundary conditions.}
\label{fig:mesh_gudget_close}
\end{figure}

\noindent A representation of the mesh is reported in Fig. \ref{fig:mesh_gudget_far} and Fig. \ref{fig:mesh_gudget_close}. The boundary patches are coloured based on the boundary condition applied. In red, the inlet is pictured, with orange we represent the outlet, while green identifies no penetration. Finally, the no slip boundary condition is identified in blue. Please note that some surfaces are not solid coloured, but only wireframed, for visualisation purposed. In the figures, it can clearly be seen the refinement close to the model surface due to the prison layer cells mentioned above.


\noindent As the mesh is relatively coarse, FGMRES was used as a linear solver for the local solution. Indeed, we are not interested in detailed aerodynamic performances, but rather on macroscopic aeroelastic properties. The latter usually require coarser mesh and this was exploited to speed up the computations. Further, a local CFL number of 40 was used for the same purpose.

\noindent The physical time step size is 0.5 ms and 600 time steps are performed, in order to assess the stability. Of this 600 time steps, the first 120 are actually purely aerodynamic. Indeed, we first initialise the correct aerodynamics, and only later we let the coupled system evolve. This is done in order to avoid spurious overshoots in the structural solution, at the very beginning of the simulation.

\begin{figure}[hp]
\centering
\includegraphics[width=0.47\textwidth]{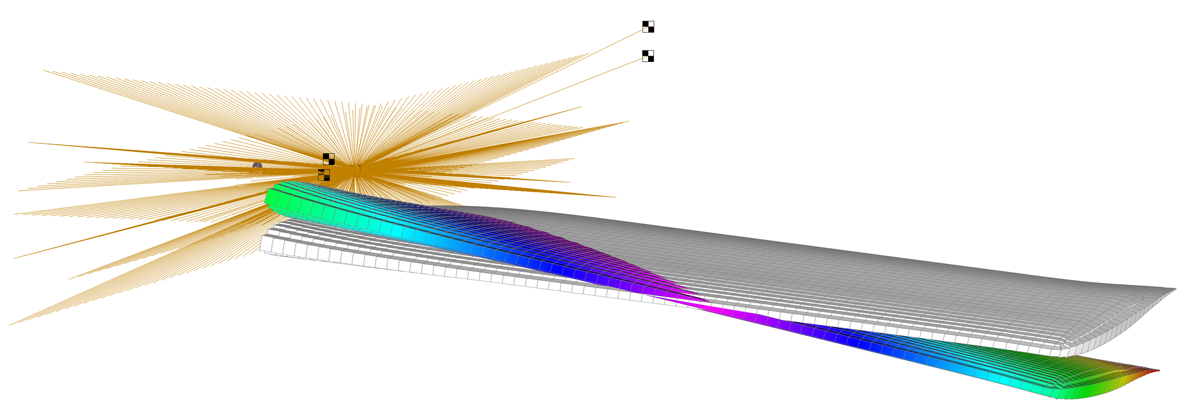}\quad\quad
\includegraphics[width=0.47\textwidth]{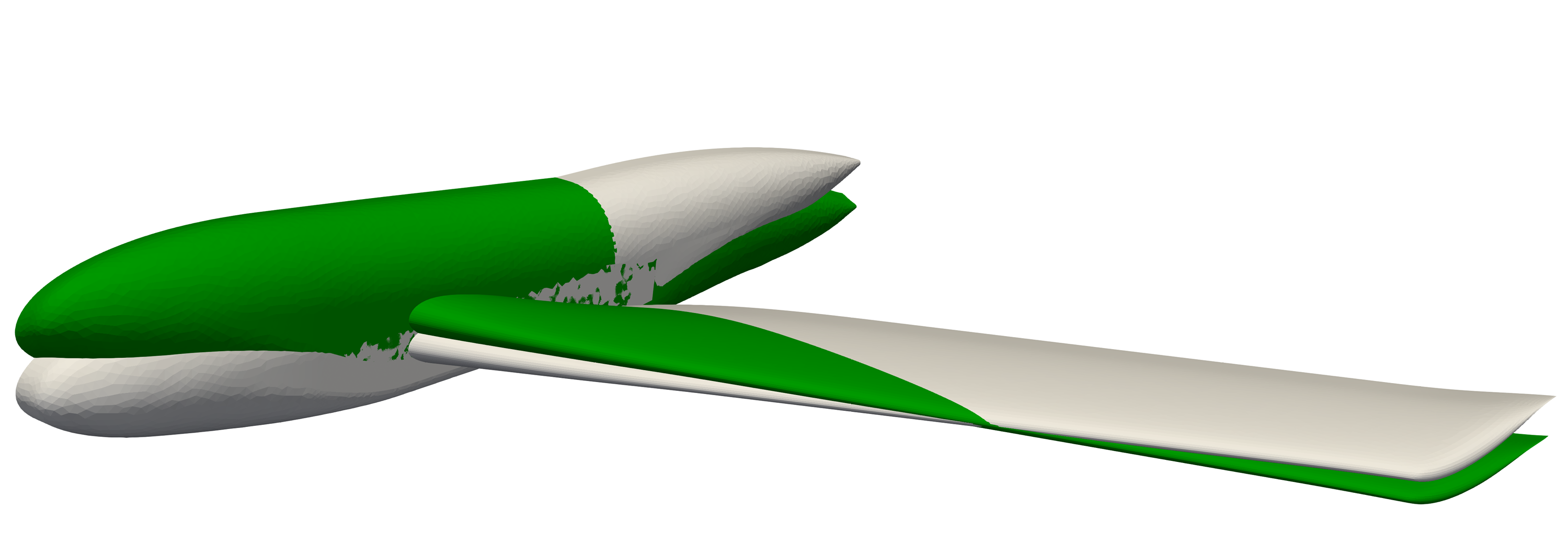}
\caption{Comparison between pitch modes at the structural and aerodynamic side ($20.1\:Hz$).}
\label{fig:modes_pitch}
\end{figure}

\begin{figure}[htp]
\centering
\includegraphics[width=0.47\textwidth]{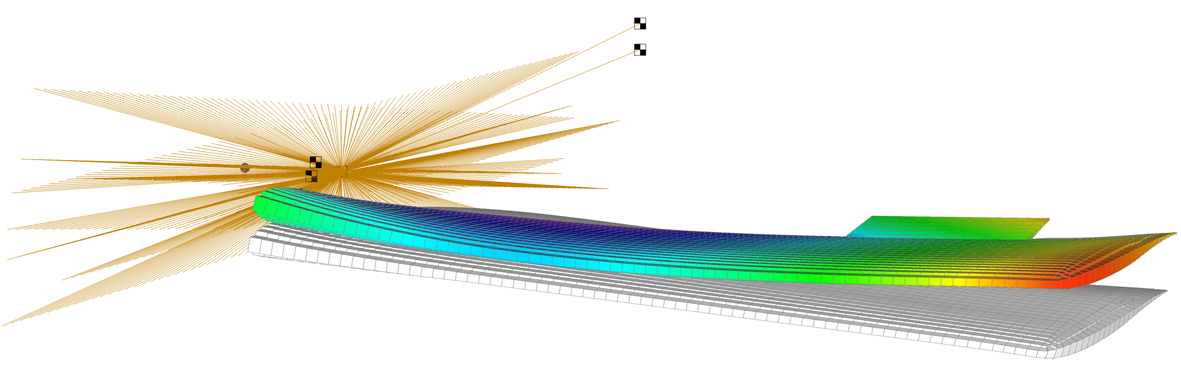}\quad\quad
\includegraphics[width=0.47\textwidth]{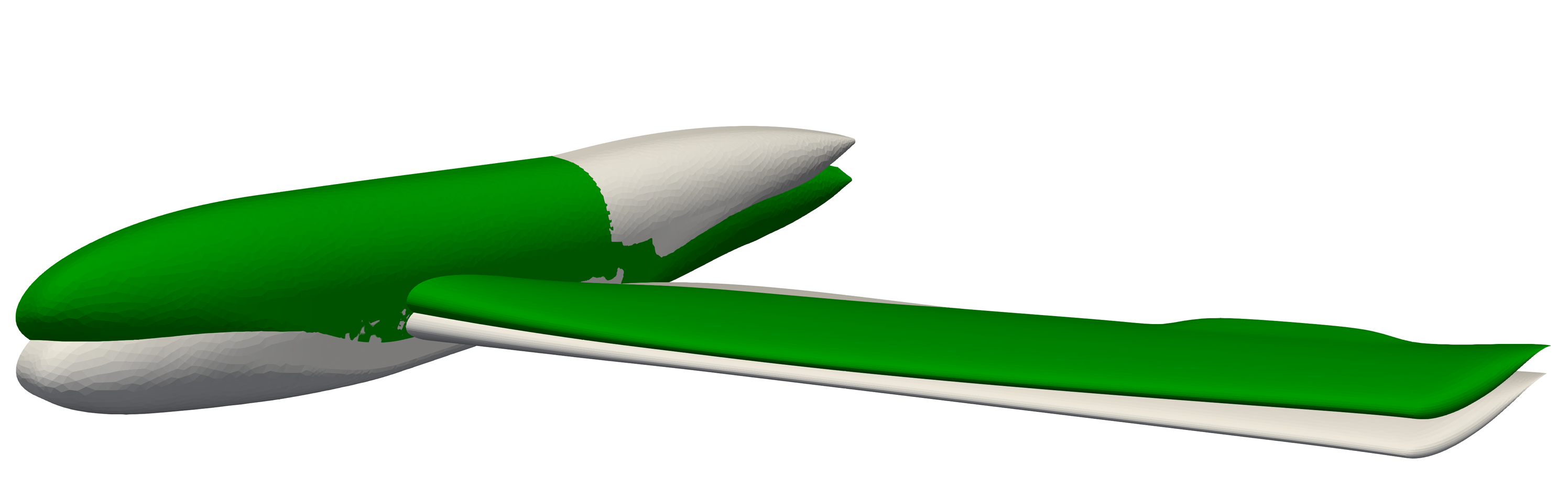}
\caption{Comparison between first bending modes at the structural and aerodynamic side ($63.1\:Hz$).}
\label{fig:modes_1st}
\end{figure}

\begin{figure}[htp]
\centering
\includegraphics[width=0.47\textwidth]{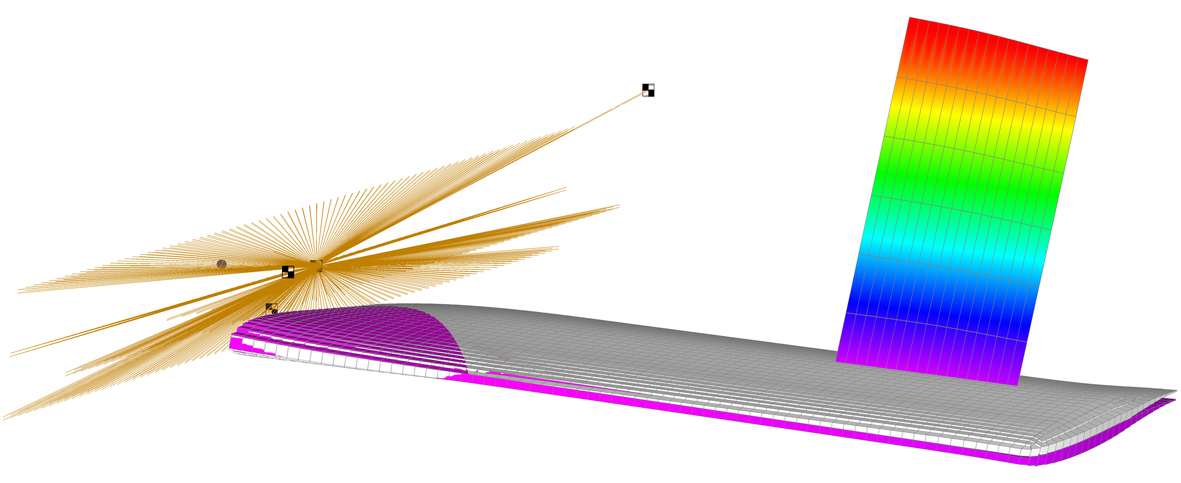}\quad\quad
\includegraphics[width=0.47\textwidth]{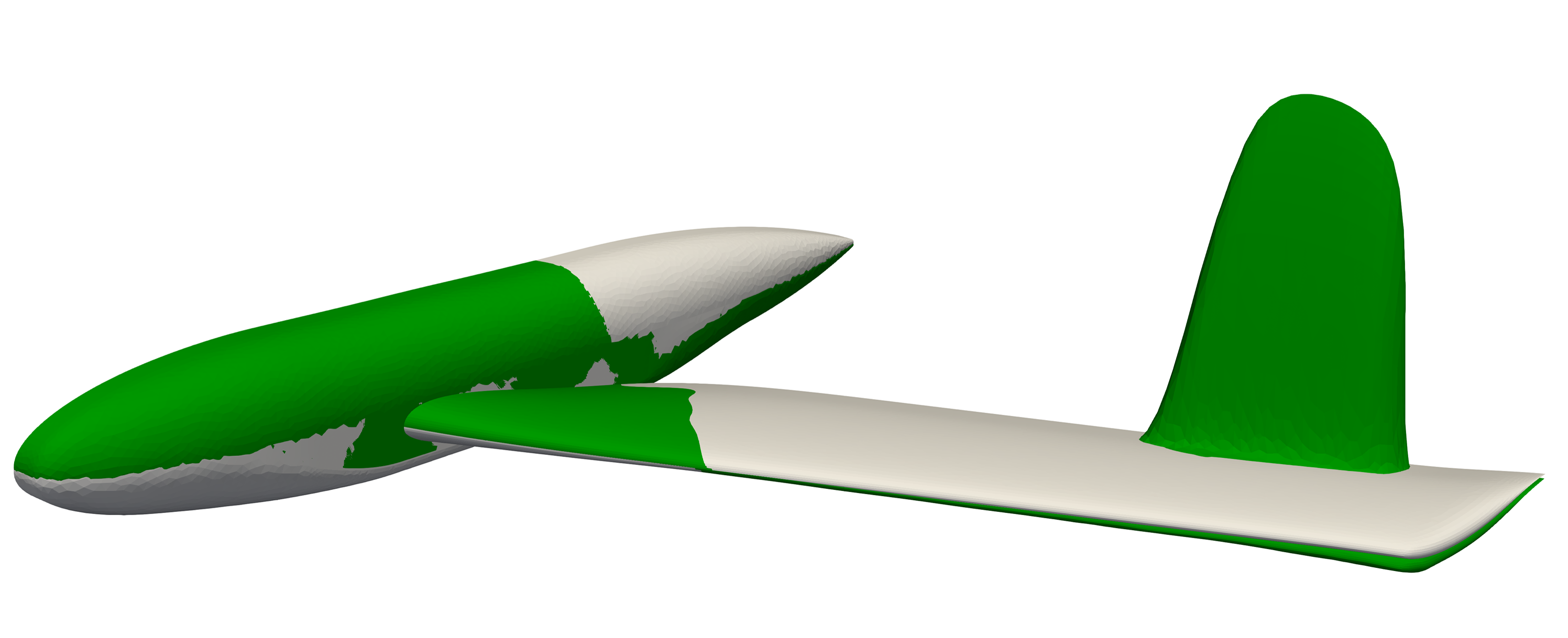}
\caption{Comparison between aileron modes at the structural and aerodynamic side ($120.7\:Hz$).}
\label{fig:modes_2nd}
\end{figure}

\begin{figure}[htp]
\centering
\includegraphics[width=0.47\textwidth]{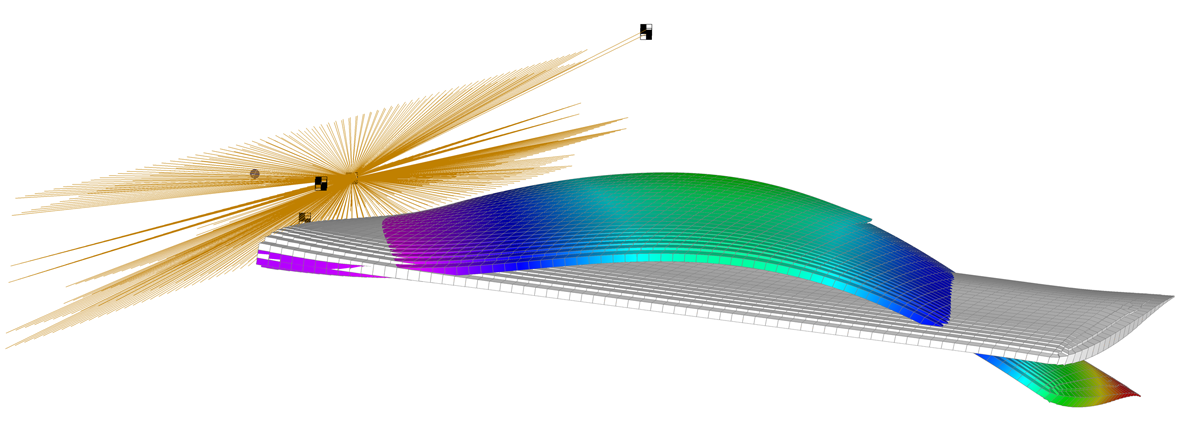}\quad\quad
\includegraphics[width=0.47\textwidth]{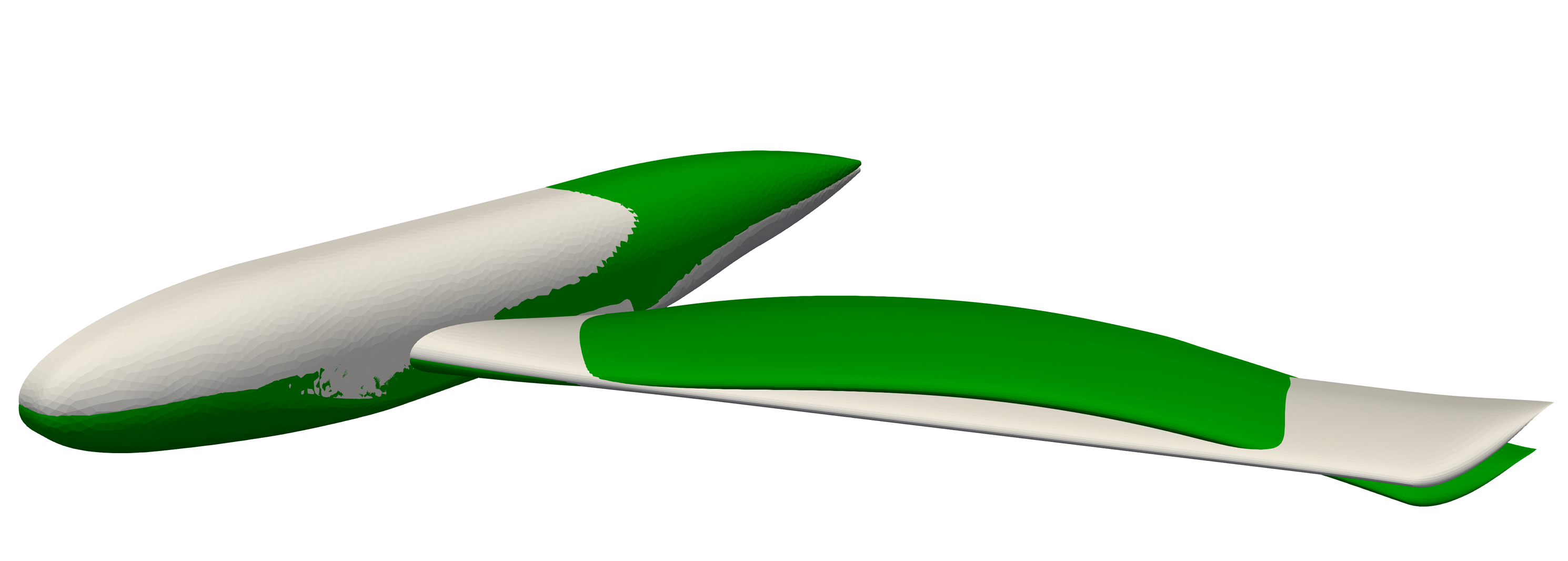}
\caption{Comparison between second bending modes at the structural and aerodynamic side ($252.4\:Hz$).}
\label{fig:modes_3rd}
\end{figure}

\begin{figure}[htp]
\centering
\includegraphics[width=0.47\textwidth]{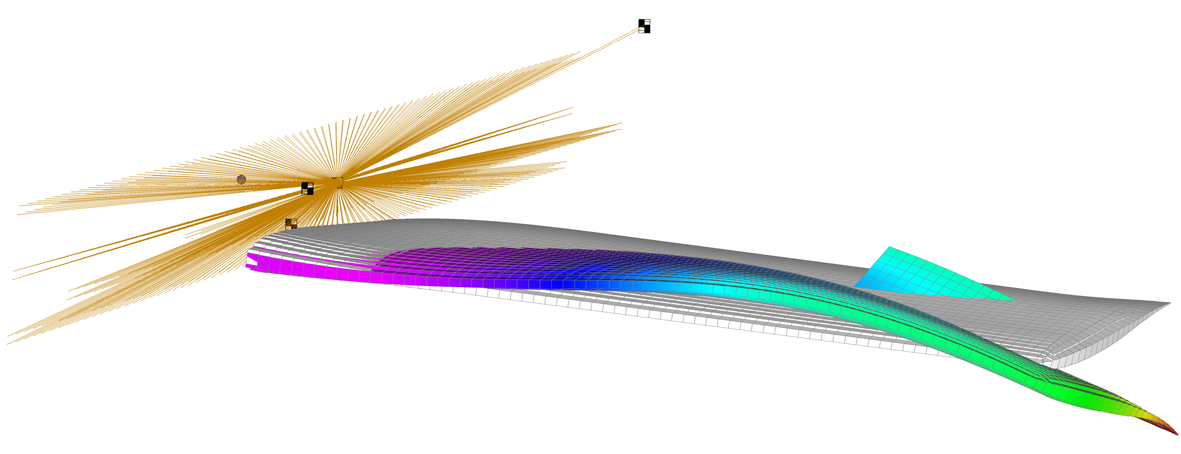}\quad\quad
\includegraphics[width=0.47\textwidth]{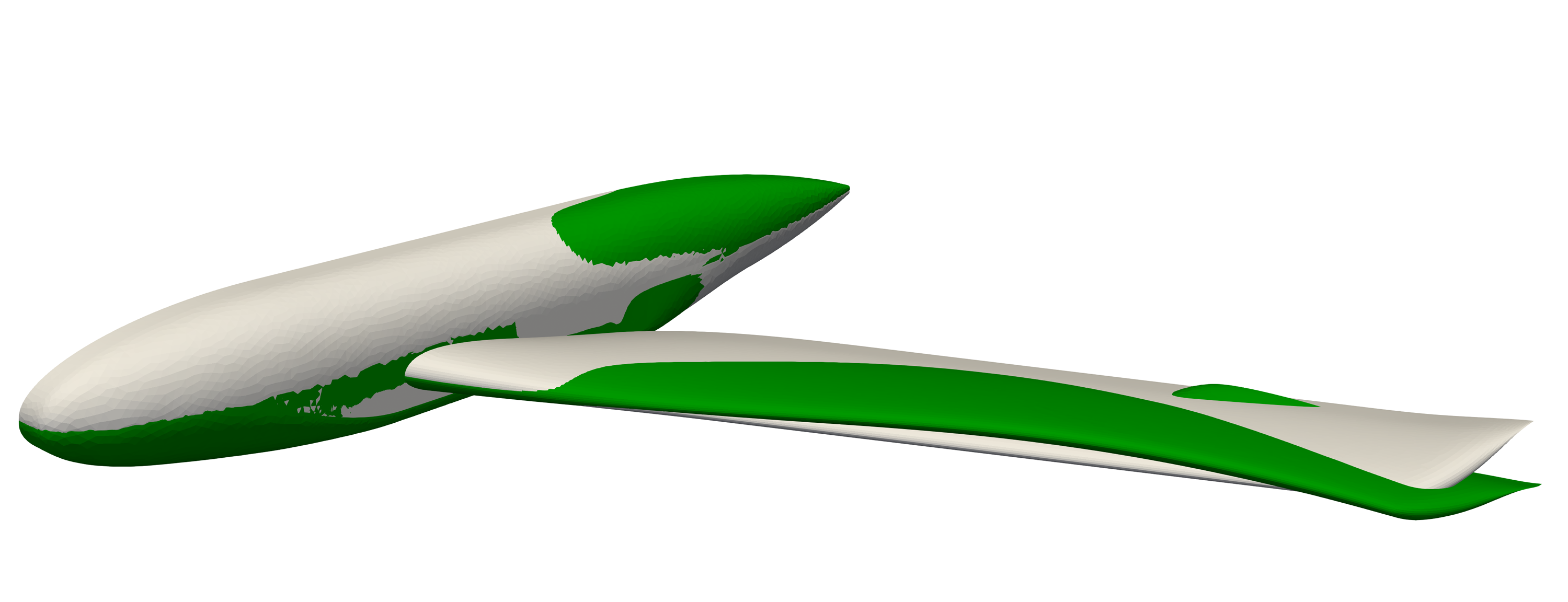}
\caption{Comparison between bending/torsional modes at the structural and aerodynamic side ($352\:Hz$).}
\label{fig:modes_4th}
\end{figure}

\begin{figure}[htp]
\centering
\includegraphics[width=0.47\textwidth]{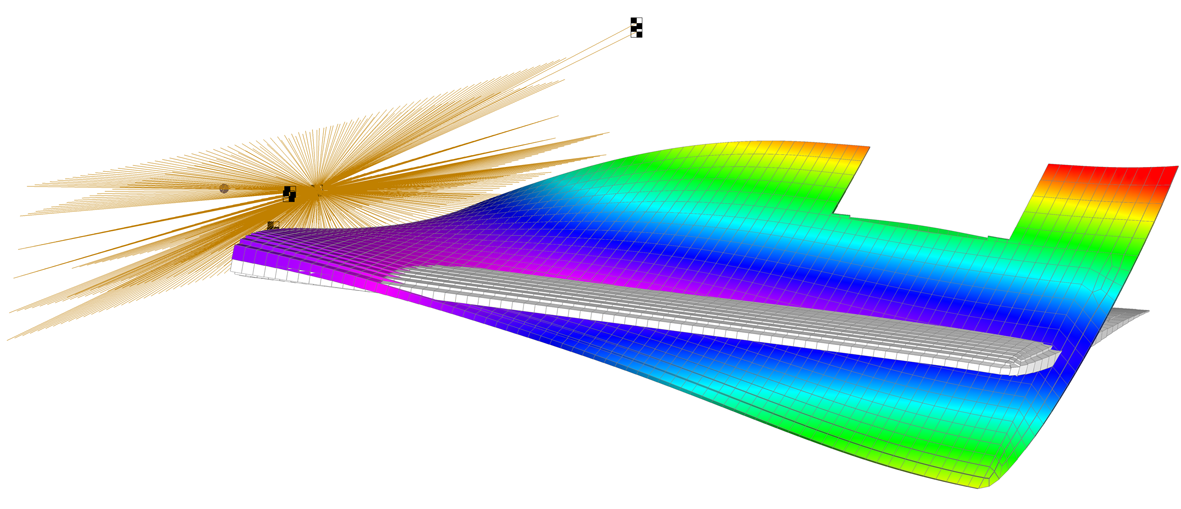}\quad\quad
\includegraphics[width=0.47\textwidth]{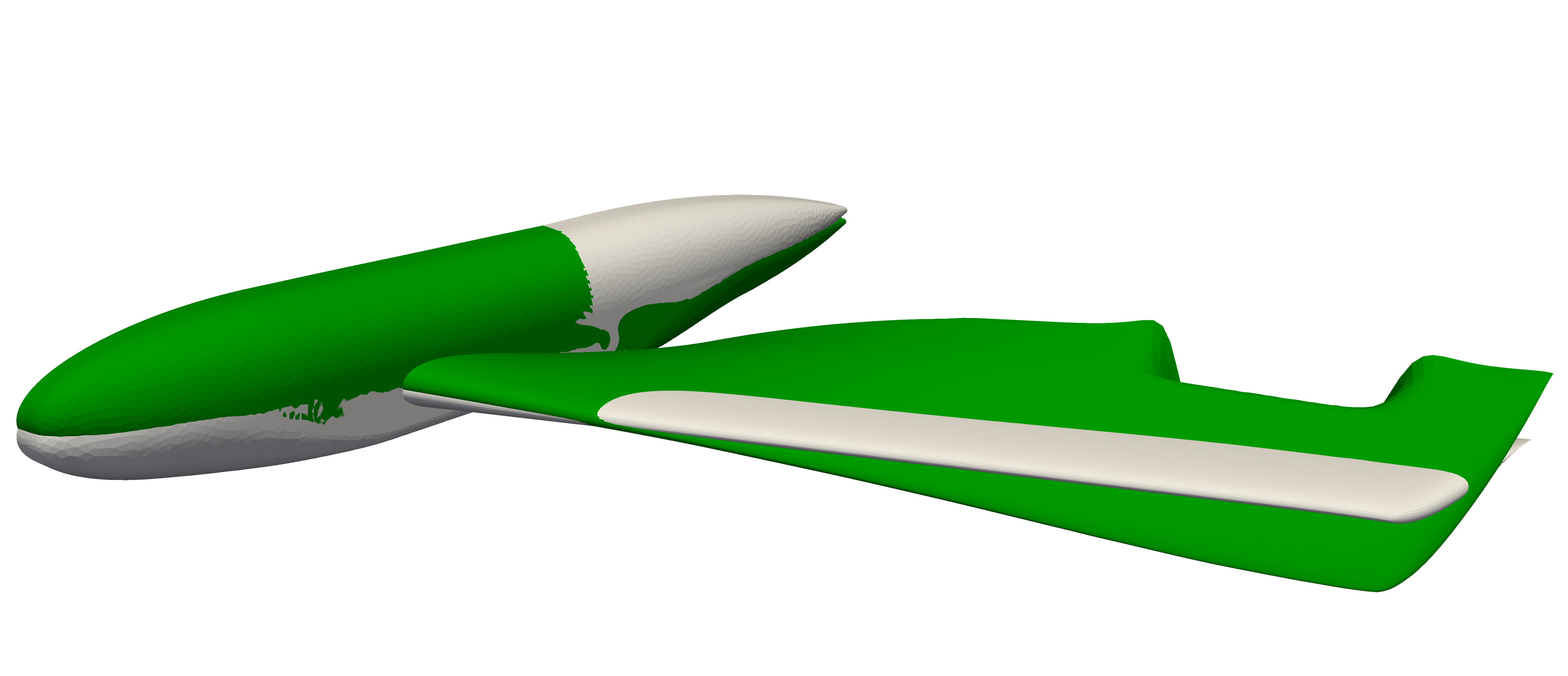}
\caption{Comparison between torsional modes at the structural and aerodynamic side ($459.5\:Hz$).}
\label{fig:modes_5th}
\end{figure}

In Fig. \ref{fig:modes_pitch} through \ref{fig:modes_5th}, the modes used for the simulation are reported.
On the left, the modes are plotted on the structural mesh. On the right, the same modes are also reported after the RBF mapping from structural mesh to aerodynamic one. It can be seen how the shapes match nicely, proving the quality of our methodology.


In the following results, different rotational axis positions are considered. The rotational axis is originally placed at the quarter Mean Aerodynamic Chord (MAC). The stability at the nominal operating condition of Mach 0.82 is first assessed. Then, in order to verify the sensitivity of this stability to changes in the rotational axis position, the latter is moved of $\pm 10\% \, \mathrm{MAC}$ from the original position and the stability analyses are repeated. The center of mass of the shaft sustaining the model is also moved to correctly represent the new mass distribution.

\subsubsection{Rigid pitch stability analysis}

As stated above, the model has different elastic modes, but can also pitch with respect to the wind tunnel. The pitch is restrained with an elastic system, introducing in the system a characteristic frequency related to this body-wide movement. This frequency is the lowest one in the system itself, thus it is of paramount importance to understand whether it will be stable or not. Indeed, the gust excitation provided to the model will mostly influence this trajectory, and the model must be dynamically stable in all cases.

\begin{figure}[ht]
\centering
\includegraphics[width=0.495\textwidth]{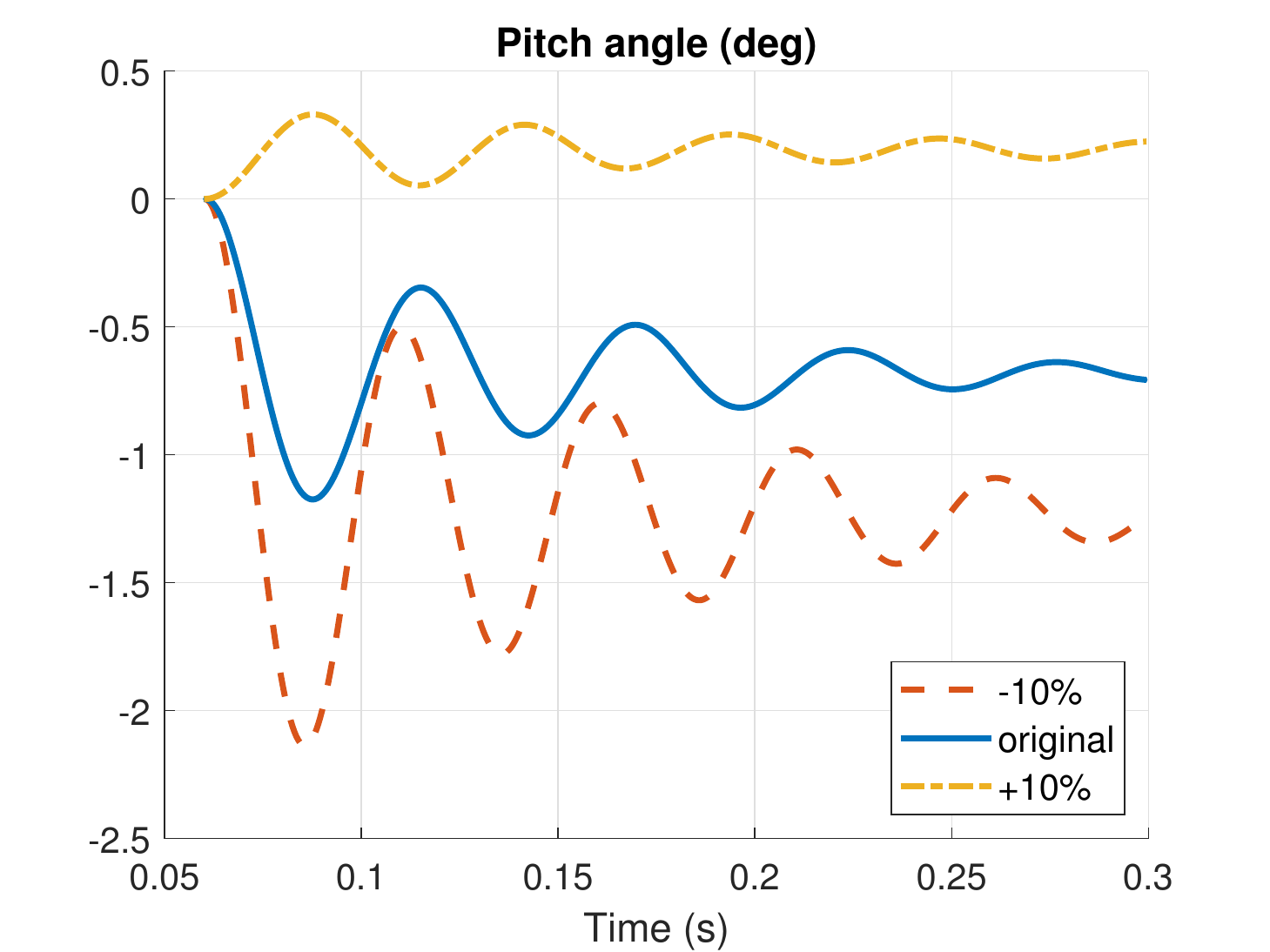}
\includegraphics[width=0.495\textwidth]{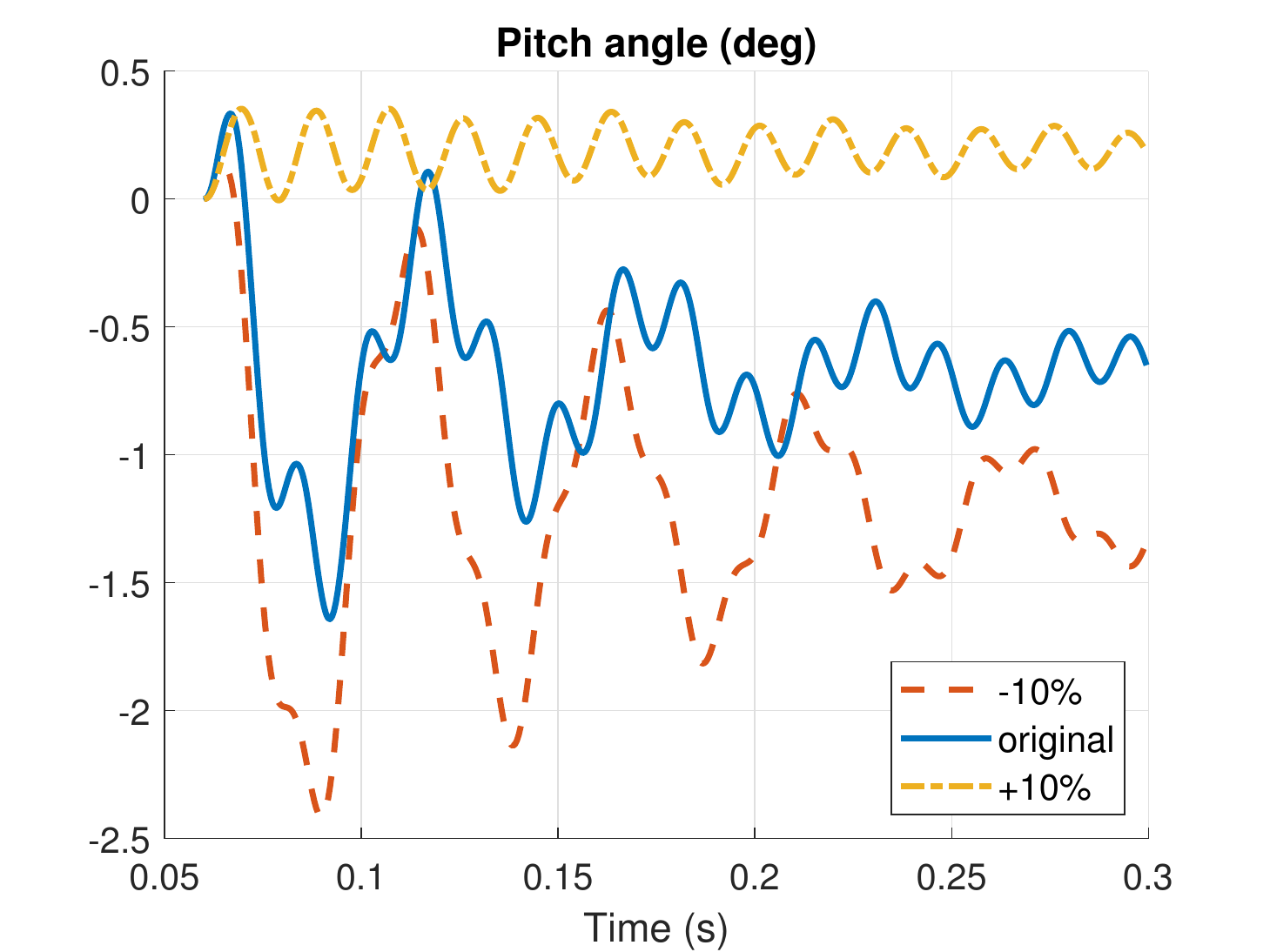}
\caption{Time histories of the GUDGET model pitch degree of freedom, for different rotational axis position, considering rigid (left) and fully flexible (right) model.}
\label{fig:rigidvsflexible}
\end{figure}

\noindent For this reason, the first set of analyses concentrated on this movement only. The model was kept rigid, except for the first mode which represents the pitch itself. In Fig. \ref{fig:rigidvsflexible} left, time histories of the pitch, for the different positions, are reported.

At the beginning of the simulation the system is in the undeformed configuration, which has no angle of attack with respect to the wind. However, this position is not the equilibrium position and this is sufficient to excite oscillations around the pitch axis. Depending on the relative position between the center of rotation and the center of pressure, the equilibrium point may be characterised by a positive or negative angle of attack. It can clearly be seen that the system, in these aerodynamic conditions, is stable. Few cycles are required to damp out the oscillations and reach the asymptotic value. 

When we shift towards the leading edge the rotational axis, a larger negative pitch moment is created, meaning larger oscillations at the beginning of the simulation, and a larger negative equilibrium value. Further, it can be noted that, when the rotational axis is placed at 35\% of the MAC, the equilibrium point shows a positive angle of attack. It is then clear that, for an intermediate position between 35\% and 25\% there must be a point where the equilibrium is exactly at zero angle of attack. This is important to note as, in the proximity of this position for the rotational axis, we may find the minimum required torque for the pitch actuation system. It is also to be expected that, if the rotational center is moved further in the positive x direction, an unstable condition would be found. However, in the range of our interest, this was not the case.

\begin{figure}[ht]
\centering
\includegraphics[width=0.49\textwidth]{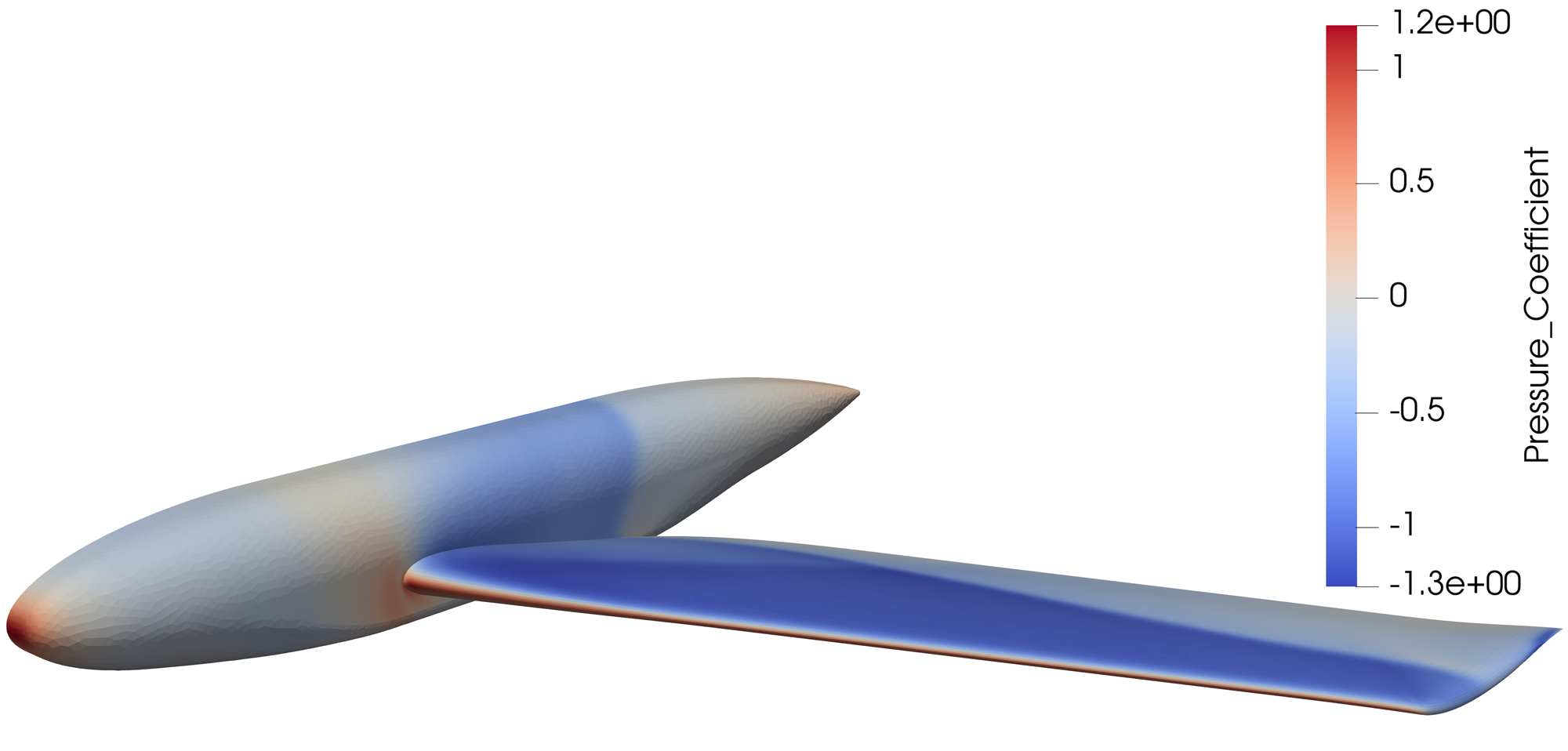}
\includegraphics[width=0.49\textwidth]{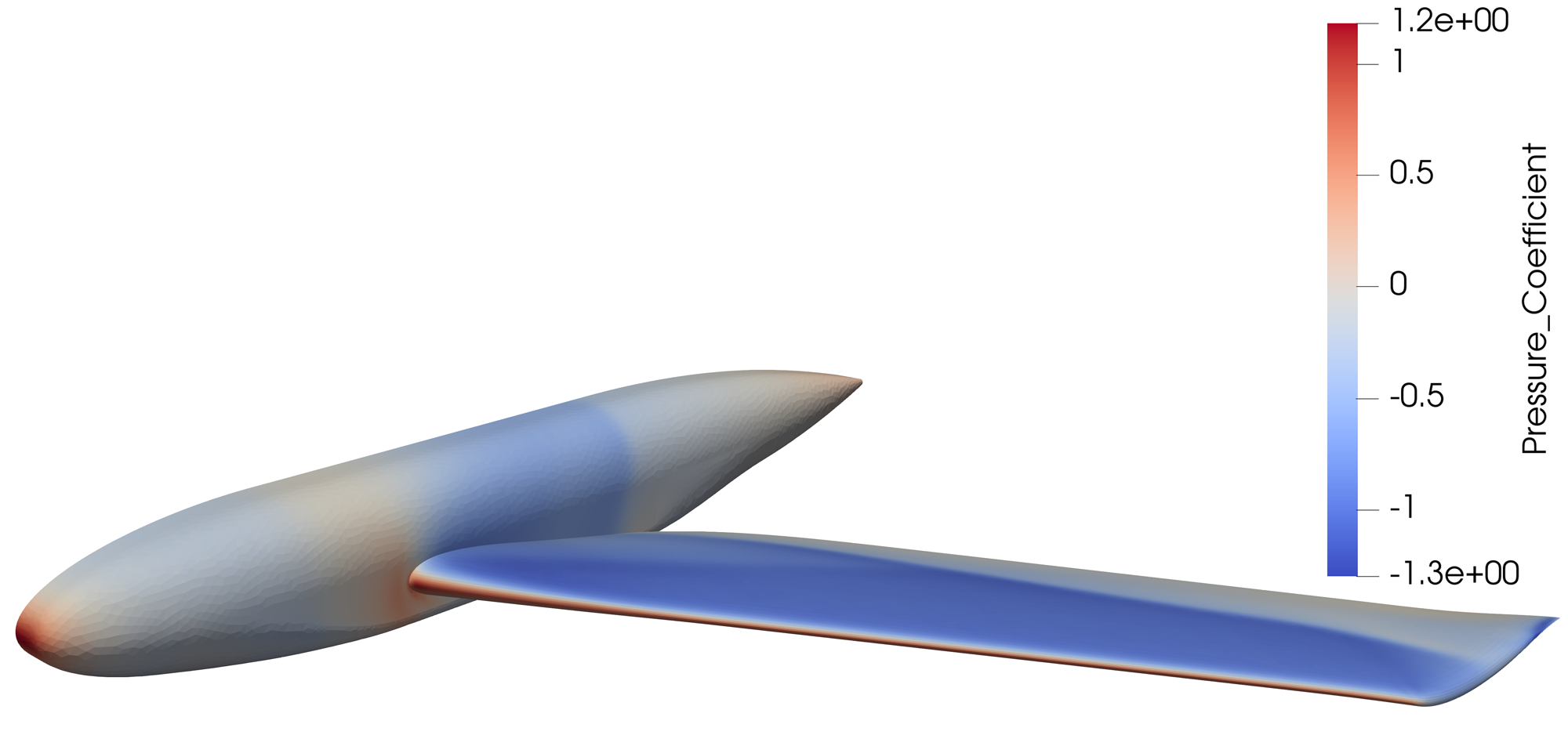}
\caption{Pressure coefficients at the time instants corresponding to the maximum (left) and minimum (right) angle of attack.}
\label{fig:Cp_compare}
\end{figure}

The stability of the system is probably due to the shock. This can be seen in Fig. \ref{fig:Cp_compare}. Here the evolution of the pressure coefficient is reported, for two time moments, for the original position of the rotational axis. The first time instant corresponds to the maximum peak in angle of attack, which is also the initial condition, the last one to the minimum peak. It can be seen the movement of the shock towards the leading edge and its increase in strength, as the model pitches up. However, this mostly occur at the tip, due to the twisted design of the wing.

\noindent Due to this effect, together with the sweeping of the wing, the overall center of pressure shifts back due to a positive pitching, stabilising the model.

\subsubsection{Flexible body and free pitch stability analysis}

After the confirmation that the pitch rotation was stable in the operating condition, the full flexibility of the system was taken into account. In Fig. \ref{fig:rigidvsflexible} right, the time histories of the pitch degree of freedom are again reported, but this time contributions at different frequencies can be noted. Indeed, together with the pitch rotation, which has a frequency of 20 Hz, higher frequencies appear in the solution. Mainly, the bending frequency at approximately 60 Hz. Again, in all cases, the system is stable. However, it can be seen that the contributions at higher frequencies are less damped than the pitch mode. Especially when the rotational axis is displaced of +10\%, the bending mode is only slightly reduced. It is then expected, at higher speeds, to encounter possible dynamic instabilities.

\section{Conclusions}\label{sec:conclusions}

In the present paper, an update and extension of the Python--based FSI framework embedded in SU2, for general aeroelastic studies, has been presented. The software, the first version of which derived from CUPyDO, couples a well known, actively developed, finite volume code, SU2, with different structural solvers, via a python interface. Thanks to the improved version, higher performances can be obtained, and an easier inclusion of new structural solvers is possible. Further, the new interface allows for four possible solutions; steady and unsteady simulations, but also imposed motion responses, covering a wide spectrum of applications. A new native structural solver allows for the direct solution of FSI problems without the need for other external structural solvers. Thanks to the very high level of the programming language, extensions and modifications are easily included.

\noindent The native solver is designed to solve a set of structural equations of motion coming from a Nastran--like model. This allows for an integration of the present tool also in industrial or standardised workflows.

The implementation has been extensively tested on various test cases, of increasing complexity. The first test was a standard pitching-plunging airfoil, operating in subsonic conditions, for which analytical results are available. In the second case, a transonic application was considered, with a three-dimensional wing, for which experimental values are present. The last test was performed on a fully flexible half-plane wind tunnel model, operating in the deep transonic regime. This involved highly nonlinear aerodynamic phenomena and demonstrated the great capabilities of the approach.

The open-source coding is another important feature of the present effort, as this increases significantly the spectrum of the possible users. The hope of the authors is that more and more practitioners will tend towards high-fidelity aeroelasticity thanks this work.

In the future, extensions are planned to allow for an automatic identification of aerodynamics thanks to the possibility of imposing the structural motion. Different techniques can then be implemented that will eventually provide with a state space approximation of the aerodynamics. This will fill the gap between the efficiency of the low-fidelity method DLM and the computationally intensity of the high-fidelity method used in this work. Indeed, the user will be provided with the opportunity to run only few full simulations, and reuse the linearised solutions afterwards.

\section*{Declaration of competing interest}

\noindent The authors declare that they have no known competing financial interests or personal relationships that could have appeared to influence the work reported in this paper.

\section*{Acknowledgments}

\noindent This project has received funding from the Clean Sky 2 Joint Undertaking under the European Union's Horizon 2020 research and innovation programme under grant agreement No 831802.

\bibliographystyle{abbrv}
\bibliography{arxiv_fsicfd.bib}

\end{document}